\documentclass[12pt]{article}
\pdfoutput=1

\usepackage{array} 
\usepackage{amssymb}
\usepackage{graphics,graphpap}
\usepackage{graphicx}
\usepackage{color}
\usepackage{graphicx}% Include figure files
\usepackage{dcolumn}% Align table columns on decimal point
\usepackage{epsfig}
\usepackage{epstopdf}
\usepackage{amsmath}
\usepackage{amsfonts}
\usepackage{textcomp}
\usepackage{setspace}
\usepackage{slashed}
\usepackage{footnote}
\makesavenoteenv{table}
\usepackage{url}

\newcommand{\ra}[1]{\renewcommand{\arraystretch}{#1}}

\newcommand{\beq}{\begin{eqnarray}}
\newcommand{\eeq}{\end{eqnarray}}

\newcommand{\bmp}{\noindent\begin{minipage}{16cm}}
\newcommand{\emp}{\end{minipage}\vskip 7mm} % 7mm untightened

\def\drawbox#1#2{\hrule height#2pt
        \hbox{\vrule width#2pt height#1pt \kern#1pt
              \vrule width#2pt}
              \hrule height#2pt}

\def\Asym#1#2{\vcenter{\vbox{\drawbox{#1}{#2}
              \kern-#2pt % line up boxes
              \drawbox{#1}{#2}}}}

\def\simge{\mathrel{%
   \rlap{\raise 0.511ex \hbox{$>$}}{\lower 0.511ex \hbox{$\sim$}}}}

\def\simle{\mathrel{
   \rlap{\raise 0.511ex \hbox{$<$}}{\lower 0.511ex \hbox{$\sim$}}}}

\def\s#1{\setbox0=\hbox{$#1$}%
\rlap{\ifdim\wd0>.7em\kern.22\wd0\else\kern.1\wd0\fi /}#1}

\usepackage{cite}

\begin{document}
%%%%%%%%%%%%%%%%%%%%%%%%%%%%%%%%%%%%%%%%%%%%%%%%%%%%%%%%%%%%%%%%%%%%%%%%%%%

%%%%%%%%%%%%%%%%%%%%%%%%%%%%%%%%%%%%%%%%%%%%%%%%%%%%%%%%%%%%%%%%%%%%%%%%%%%
%%%%%%%%%%%%%%%%%%%%%%%%%%%%%%%%%%%%%%%%%%%%%%%%%%%%%%%%%%%%%%%%%%%%%%%%%%%
%\begin{titlepage}
%\title{\vspace*{-2.0cm}
%\hfill {\small MPP-2015-302}\\[20mm]
%\vspace*{-1.5cm}
%\bf\Large
%Resonantly produced sterile neutrino dark matter: Constraints from structure formation\\[5mm]\ \vspace{-1cm}}
%Constraints from structure formation on resonantly produced sterile neutrino dark matter\\[5mm]\ \vspace{-1cm}}

\begin{titlepage}
\title{
%Resonantly produced sterile neutrino dark matter: Constraints from structure formation\\[5mm]\ \vspace{-1cm}}
%Constraints from structure formation on resonantly produced sterile neutrino dark matter\\[5mm]\ \vspace{-1cm}}
Astrophysical constraints on resonantly produced sterile neutrino dark matter\\[5mm]\ \vspace{-1cm}}

\author{
Aurel Schneider\thanks{email: \tt aurel@physik.uzh.ch}\\ \\
{\normalsize \it Center for Theoretical Astrophysics and Cosmology,}\\
{\normalsize \it Institute for Computational Science, University of Z\"urich,}\\
{\normalsize \it Winterthurerstrasse 190, CH-8057, Z\"urich, Switzerland.}\\
}
\date{\today}
\maketitle
\thispagestyle{empty}

\vspace{-0.5cm}
\begin{abstract}
\noindent
Resonantly produced sterile neutrinos are considered an attractive dark matter (DM) candidate only requiring a minimal, well motivated extension to the standard model of particle physics. With a particle mass restricted to the keV range, sterile neutrinos are furthermore a prime candidate for warm DM, characterised by suppressed matter perturbations at the smallest observable scales. In this paper we take a critical look at the validity of the resonant scenario in the context of constraints from structure formation. We compare predicted and observed number of Milky-Way satellites and we introduce a new method to generalise existing Lyman-$\alpha$ limits based on thermal relic warm DM to the case of resonant sterile neutrino DM. The tightest limits come from the Lyman-$\alpha$ analysis, excluding the entire parameter space (at 2-$\sigma$ confidence level) still allowed by X-ray observations. Constraints from Milky-Way satellite counts are less stringent, leaving room for resonant sterile neutrino DM most notably around the suggested line signal at 7.1 keV.
\end{abstract}
\end{titlepage}

%%%%%%%%%%%%%%%%%%%%%%%%%
\section{\label{sec:intro}Introduction}
Despite enormous efforts in both fields particle physics and cosmology, the nature and composition of dark matter (DM) is still largely unknown. Up to date, the most popular DM candidate, the lightest stable particle from supersymmetry (SUSY), has neither be found via direct \cite{Aprile:2012nq,Akerib:2013tjd} nor indirect detection \cite{Adriani:2013uda,Ackermann:2015zua}, nor has any convincing sign of SUSY been established with particle colliders \cite{Aad:2014vma,Khachatryan:2014qwa}.

In this situation it is quite natural to search for alternative DM candidates outside of the SUSY framework. A very promising candidate is the sterile neutrino. Adding a family of sterile singlets to the neutrino sector consists of a simple and natural extension of the standard model of particle physics. Sterile neutrinos act as right-handed counterpart to the left-handed active neutrinos and they may provide a natural framework for the observed non-zero masses of the active neutrinos via the seesaw mechanism\footnote{At least three sterile singlets are required to solve both the neutrino mass and the dark matter problem \cite{Kusenko:2009up}.}.

The only coupling of sterile neutrinos with standard model particles is via flavour mixing to active neutrinos. Sterile neutrino DM is therefore most naturally produced by the {\it freeze-in} mechanism, i.e. the DM abundance is gradually build up from the active neutrinos which are part of the primordial plasma. This most minimal production mechanism has first been advocated in a paper by Dodelson \& Widrow (DW) from the early nineties \cite{Dodelson:1993je} (but see also \cite{Barbieri:1989ti}). However, the DW mechanism produces rather large particle momenta and has been shown to be in conflict with either Lyman-$\alpha$ observations or X-ray data, depending on the particle mass \cite{Seljak:2006qw,Viel:2006kd}.

In the case of a significant lepton asymmetry of the primordial plasma, sterile neutrinos can be produced resonantly, requiring smaller mixing angles and yielding significantly colder momenta. This has first been pointed out by Shi \& Fuller (SF) \cite{Shi:1998km} and is usually referred to as the \emph{resonant} or SF production mechanism. Resonantly produced sterile neutrinos are believed to evade the strong constraints from Lyman-$\alpha$ observations and are considered an attractive DM candidate \cite{Boyarsky:2009ix}. They are furthermore one of the cornerstones of the \emph{Neutrino Minimal Standard Model} ($\nu$MSM) which is a framework attempting to solve the dark matter, the baryon asymmetry, and the nonzero neutrino mass problems by adding only three additional sterile neutrinos to the standard model \cite{Asaka:2005an,Asaka:2005pn,Canetti:2012kh}. 

In terms of structure formation, there are numerous studies putting forward limits for the non-resonant scenario \cite{Seljak:2006qw,Viel:2006kd,Polisensky:2010rw,deSouza:2013hsj,Kennedy:2013uta,Inoue:2014jka,Menci:2016eww}, but only very few for resonant sterile neutrino DM~\cite{Horiuchi:2015qri,Lovell:2015psz}. The main reason for the lack of competitive constraints is the spiky nature of momentum distributions due to the resonance, which leads to a non-trivial suppression of perturbations at small scales, making a thorough statistical analysis a challenging task. 

In this paper we take a fresh look at constraints from structure formation and how they affect resonant sterile neutrino DM. We develop a method to apply existing Lyman-$\alpha$ limits, based on the non-resonant scenario, to the more general case of resonant sterile neutrino DM. Furthermore, we compute constraints from Milky-Way satellite counts, using a method which allows for arbitrarily shaped power spectra~\cite{Schneider:2013ria,Schneider:2014rda}.

The paper is structured as follows: In Sec.~\ref{sec:production} we summarise the resonant production mechanism and give examples of typical momentum distributions. Sec.~\ref{sec:Xrays} provides an overview of recent constraints from X-ray observations. In Sec.~\ref{sec:structform} we take a closer look at limits imposed by structure formation from the Lyman-$\alpha$ forest and Milky-Way satellite counts. We use these limits to provide constraints on the sterile neutrino parameter space in Sec.~\ref{sec:results}, and we conclude in Sec.~\ref{sec:conclusions}.

%%%%%%%%%%%%%%%%%%%%%%%%%
\section{\label{sec:production}Resonant sterile neutrino production}
In a minimal setup sterile neutrinos are produced via oscillation with active neutrinos (with a given vacuum mixing angle $\theta$) which are weakly interacting and therefore part of the primordial plasma. In the presence of a significant primeval lepton asymmetry, production via mixing has been shown to happen resonantly, which means that sterile neutrino DM is build up more efficiently opening up the parameter space towards smaller mixing angles \cite{Shi:1998km}. The resulting momentum distribution from resonant production is very different from a Fermi-Dirac distribution (naturally obtained for DM scenarios with thermal decoupling) often exhibiting colder average momenta. The momentum distribution of a given DM scenario affects the perturbations in the early universe via the free-streaming effect \cite{Lesgourgues:2011rh} and has therefore an influence on structure formation.

For the case of collisional dominated neutrino interactions, the production of sterile singlets can be written in terms of the Boltzmann equation \cite{Abazajian:2001nj,Kishimoto:2008ic,Venumadhav:2015pla}
\begin{equation}\label{boltzmann}
\frac{\partial}{\partial t}f_{\rm sn}(p,t)-H p \frac{\partial}{\partial p}f_{\rm sn}(p,t) = \Gamma(\nu\rightarrow \nu_s;p,t) \left[f_{\nu}(p,t)-f_{\rm sn}(p,t)\right],
\end{equation}
with the active-sterile conversion rate
\begin{equation}\label{Gamma}
\Gamma(\nu\rightarrow\nu_s;p,t) = \frac{\Gamma_{\nu}(p)\Delta^2(p)\sin^22\theta}{4\{\Delta^2(p)\sin^22\theta+D^2(p)+\left[\Delta(p)\cos2\theta - V^L-V^{T}(p)\right]^2\}}\,,
\end{equation}
where $f_{\rm sn}(p,t)$ and $f_{\nu}(p,t)$ are the sterile and active neutrino distribution functions. Here, $\Gamma_\nu(p)$ is the interaction rate (between active neutrinos and the plasma), $D(p)=\Gamma_\nu(p)/2$ the quantum damping rate, and $\Delta(p)\simeq m_{\rm sn}/2p$ the vacuum oscillation rate. The weak potential of the neutrino is split into a thermal potential $V^T(p)$ and a lepton asymmetry potential $V^L$. There is a similar equation for antineutrinos not shown here. For more details, see Refs.~\cite{Abazajian:2001nj,Abazajian:2005gj,Kishimoto:2008ic,Venumadhav:2015pla}.

In the case of zero lepton asymmetry, the potential $V^L$ vanishes and Eq.~(\ref{boltzmann}) yields a final sterile neutrino momentum distribution with a shape very close to a Fermi-Dirac distribution. This is the reason for the tight connection between thermally produced warm DM (WDM) and non-resonant sterile neutrino DM. There is a simple mass conversion between $m_{\rm WDM}$ and $m_{\rm sn}$ which allows to directly compare thermal WDM and non-resonant sterile neutrino DM in terms of structure formation (see e.g. \cite{Viel:2005qj,Merle:2015vzu,Bozek:2015bdo}). In the following sections, we interchangeably refer to the thermal relic mass $m_{\rm WDM}$ or the sterile neutrino mass $m_{\rm sn}$ depending on the context.

In the presence of a primordial lepton asymmetry, the lepton asymmetry potential $V^L$ is nonzero and the resonance condition
\begin{equation}
\Delta(p_{\rm res})\cos2\theta - V^L-V^{T}(p_{\rm res}) = 0
\end{equation}
is satisfied for a given momentum $p_{\rm res}$. As a consequence, the source term of Eq.~(\ref{boltzmann}) increases dramatically, leading to instantaneous production of a large number of sterile neutrinos.

The potentials $V^L$ and $V^T$ depend on the interaction between active neutrinos and the plasma. Detailed investigations of the redistribution of lepton asymmetries and the neutrino opacities are discussed in \cite{Asaka:2006rw,Ghiglieri:2015jua,Venumadhav:2015pla}.

In this paper we use the code {\tt sterile-dm} from Ref.~\cite{Venumadhav:2015pla} to calculate resonant sterile neutrino momentum distributions. The code is restricted to oscillations with a single active neutrino species (the muon neutrino) and assumes an initial lepton asymmetry for the muon flavour. It is furthermore based on the semi-classical Boltzmann approach (i.e. Eq.~\ref{boltzmann}) which holds if the neutrino interactions are collision-dominated. We refer the reader to \cite{Venumadhav:2015pla} for a discussion of the various subtleties and possible shortcomings of the calculation.

\begin{figure}[!ht]
\center{
\includegraphics[width=.99\textwidth,trim={0cm 0cm 0cm 0cm}]{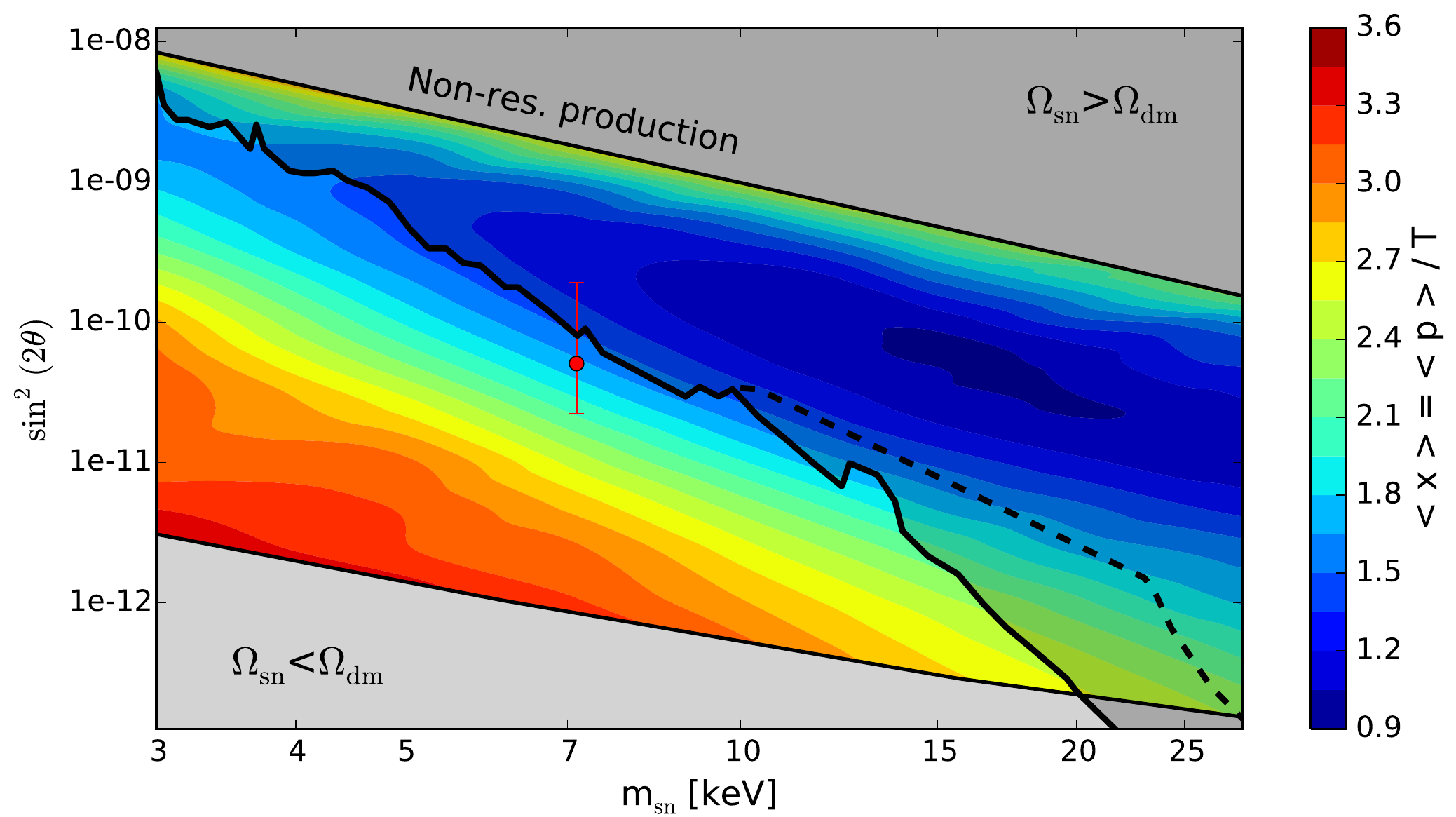}
\caption{\label{fig:xmean}Sterile neutrino mass versus mixing angle from resonant production. The parameter space is delimited by an upper and lower thin black line, corresponding to production with zero (non-resonant production) and maximum lepton asymmetry. X-ray constraints from {\it Suzaku} \cite{Tamura:2014mta,Sekiya:2015jsa} are given as thick black line (with shaded area indication the excluded region), while the dashed line corresponds to a more conservative limit from~\cite{Canetti:2012kh,Ng:2015gfa}. The colormap illustrates the mean momenta from resonant production (divided by the photon temperature $T$). The tentative line signal from Refs.~\cite{Bulbul:2014sua,Boyarsky:2014jta} is indicated by the red symbol at 7.1 keV.}}
\end{figure}

Resonant sterile neutrino DM scenarios are usually parametrised with respect to the particle mass ($m_{\rm sn}$) and the mixing angle (given by $\sin^22\theta$). The initial lepton asymmetry is then set accordingly to obtain the correct DM abundance \cite{Kishimoto:2008ic}.

In Fig.~\ref{fig:xmean} we plot the available parameter space which is restricted from all sides. The upper thin black line shows the non-resonant production, above which the sterile neutrino budget would overclose the universe (upper grey area). The lower thin black line corresponds to resonant production of maximum lepton asymmetry allowed by the $\nu$MSM model \cite{Shaposhnikov:2008pf,Laine:2008pg,Canetti:2012kh}, below which not enough sterile neutrinos are produced to make up the total DM budget (lower grey area). Form the left the parameter space is fundamentally restricted by the Tremain-Gunn bound \cite{Tremaine:1979we,Boyarsky:2008ju} (at $m_{\rm sn}\sim 1$ keV, outside of the plot) and from the right it is limited by the non-detection of emission lines from X-ray observations (solid and dashed thick black lines representing a tight and a more conservative estimate, see Sec.~\ref{sec:Xrays} for more details). %We have furthermore added the suggested X-ray line observation \cite{Bulbul:2014sua,Boyarsky:2014jta} for reference.

The colour map of Fig.~\ref{fig:xmean} shows the average momenta (divided by the photon temperature) from the momentum distributions calculated with {\tt sterile-dm} (interpolating between values on a grid covering the parameter space). As expected, the resonant production yields significantly cooler spectra than what is obtained by the non-resonant (DW) mechanism. This is, however, only true for parts of the parameter space (blue areas), while other parts are significantly warmer (red areas at very low mixing angles). The coolest spectra are found roughly one order of magnitude below the line of non-resonant production. This is in agreement with estimates from Ref.~\cite{Boyarsky:2009ix}.

\begin{figure}[tbp]
\center{
\includegraphics[width=.32\textwidth,trim={1.2cm 1.1cm 3.4cm 0cm}]{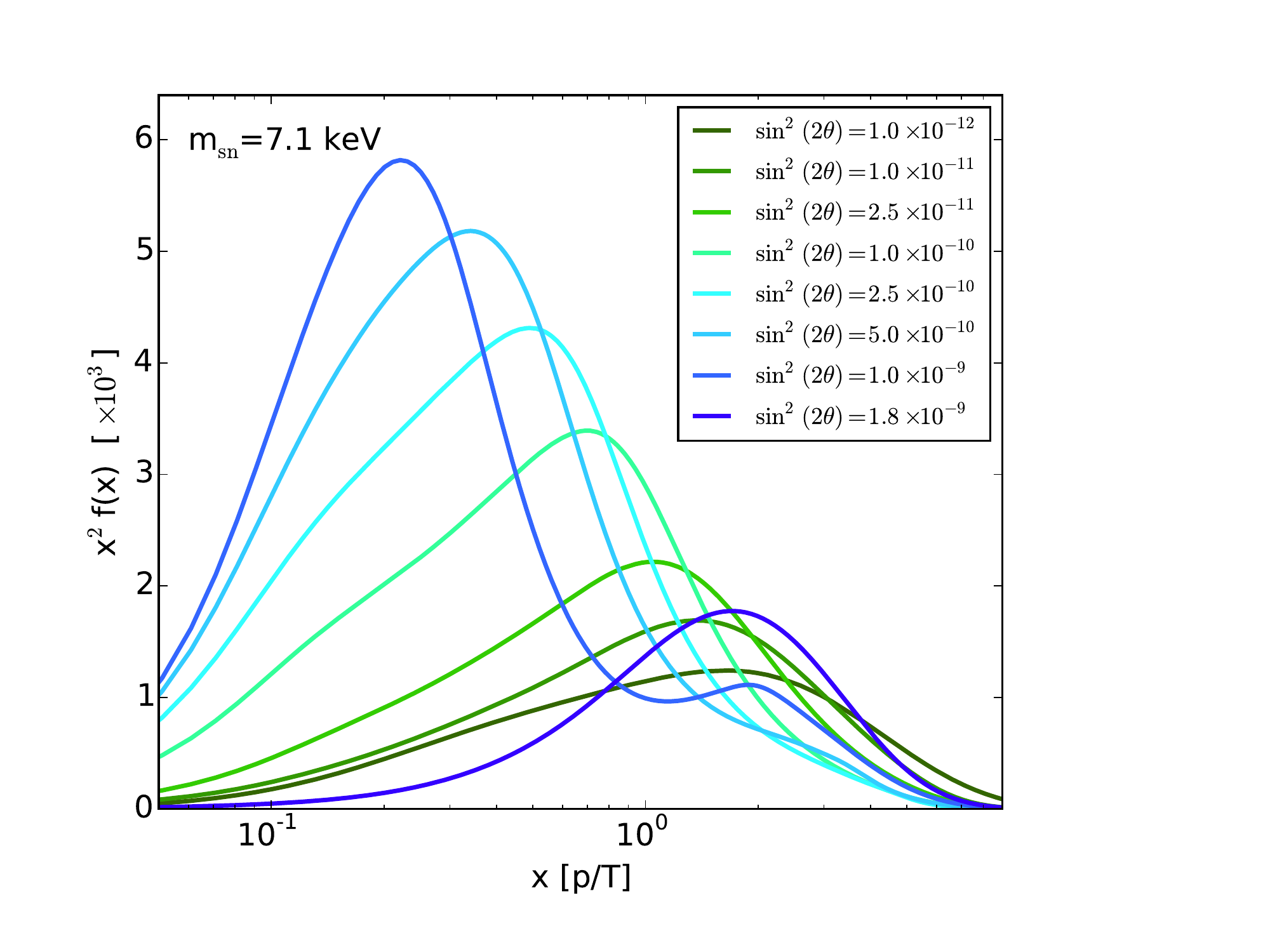}
\includegraphics[width=.32\textwidth,trim={1.2cm 1.1cm 3.4cm 0cm}]{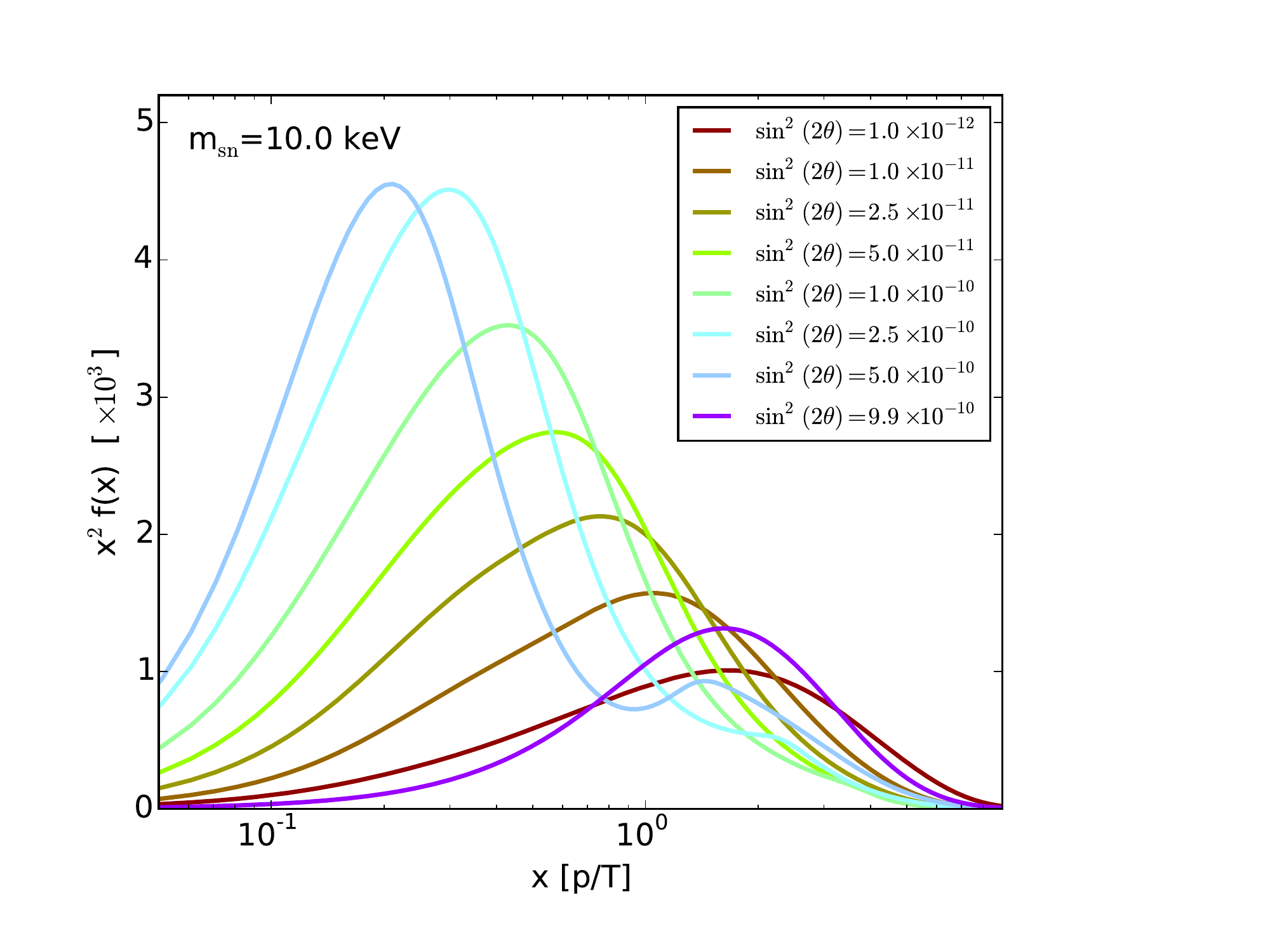}
\includegraphics[width=.32\textwidth,trim={1.2cm 1.1cm 3.4cm 0cm}]{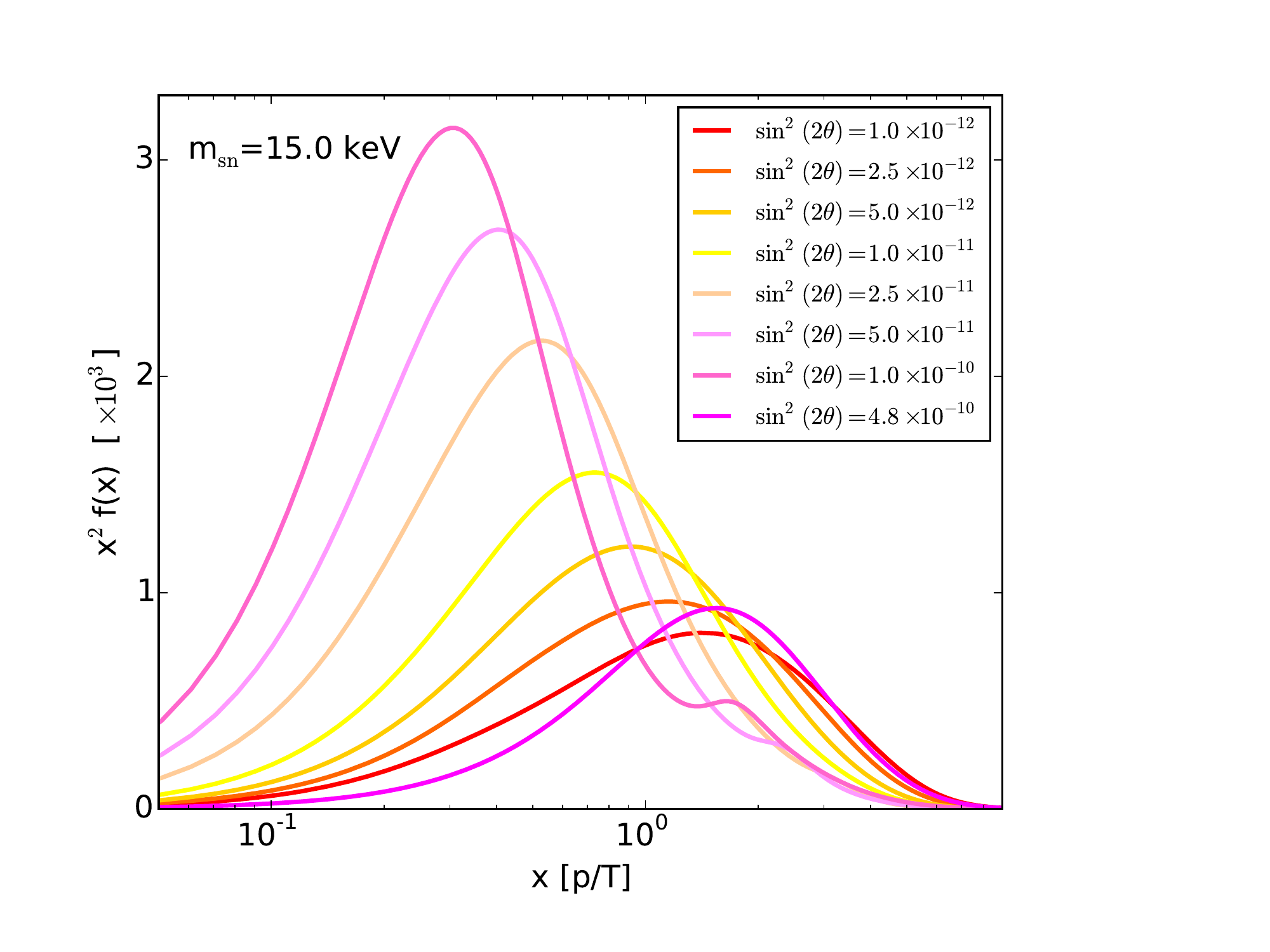}
\caption{\label{fig:dis}Momentum distribution functions of resonantly produced sterile neutrinos. From left to right: increasing values for the sterile neutrino mass. Different colours indicate different mixing angles. The model with the largest mixing angle in each panel corresponds to the non-resonant (DW) scenario.}}
\end{figure}

The mean momentum gives a first estimate of how a given scenario affects structure formation. However, a detailed investigation of the free-streaming effect requires knowledge of the full momentum distribution. This is especially true in the case of an efficient resonance where the distribution may be strongly distorted or even double-peaked. In this regime it is not advisable to estimate the free-streaming behaviour from the mean momentum alone.

In Fig.~\ref{fig:dis} we show the full momentum distributions of selected scenarios representing three different values for the particle mass (left: $m_{\rm sn}=7.1$ keV; centre: $m_{\rm sn}=10$ keV; right: $m_{\rm sn}=15$ keV) and various mixing angles (colours). The model with the largest mixing angle in every panel corresponds to the non-resonant scenario and has a momentum distribution with a shape very close to the Fermi-Dirac distribution. Decreasing the mixing angles (i.e. increasing the lepton asymmetry) first leads to a shift of the distributions towards colder momenta until a turnover point where the distributions get warmer again. The model with the smallest mixing angle in each panel of Fig.~\ref{fig:dis} has a momentum distribution which is slightly warmer than the corresponding non-resonant distribution. This behaviour confirms the results of the average momenta shown in Fig.~\ref{fig:xmean}.

There are several possible shortcomings of the calculation performed with {\tt sterile-dm}, which we now briefly discuss. First of all, the code is restricted to a single sterile singlet and only considers oscillation with the muon-neutrino ignoring mixing with other flavours\footnote{The authors of Ref.~\cite{Ghiglieri:2015jua} showed that very similar distributions may be obtained for other flavour mixings, at least for a particle mass around $m_{\rm sn}=7.1$ keV.}. Second, the calculation is based on the semi-classical Boltzmann formalism which is only accurate if collisions dominate the neutrino interactions, i.e. if $\Delta(T)\sin^22\theta/D(T)<1$ \cite{Venumadhav:2015pla}. We have checked that this is the case for all relevant parts of the parameter space. Further possible error sources are uncertainties in the calculations of neutrino opacities and the lepton asymmetry evolution during the quark-hadron transition. However, they do not seem to affect the sterile neutrino momentum distributions at a significant level (as shown in Ref.~\cite{Venumadhav:2015pla}).

%%%%%%%%%%%%%%%%%%%%%%%%%%%%
\section{\label{sec:Xrays}X-ray observations}
The sterile neutrino is expected to decay radiatively into an active neutrino $\nu$ and a photon $\gamma$ with energy $E_{\gamma}=m_{\rm sn}/2$. This leads to a line-signal which can be searched for with X-ray surveys. The decay rate \cite{Palazzo:2007gz}
\begin{equation}
\Gamma_{\nu_s\rightarrow\gamma\nu}\simeq 1.38\times 10^{-22} \sin^22\theta\left(\frac{m_{\rm sn}}{\rm keV}\right)^5\,s^{-1}
\end{equation}
strongly depends on $m_{\rm sn}$, effectively ruling out sterile neutrinos with mass significantly above the keV range as DM candidates. The expected 
flux from X-ray radiation is given by
\begin{equation}
F_{\gamma}=\frac{\Gamma_{\nu_s\rightarrow\gamma\nu}\,\Omega_{\rm fov}}{8\pi}\int_{\rm los}dx\,\rho_{\rm DM}(x)
\end{equation}
where $\Omega_{\rm fov}$ is the solid angle covering the instrumental field-of-view and the dark matter density $\rho_{\rm DM}$ is integrated over the entire line-of-sight (los). Hence, there is a direct relation between X-ray flux, mixing angle, and sterile neutrino mass which can be used to either detect sterile neutrino DM or to constrain its parameter space.

\subsection{The 3.55 keV line signal}
Recently, two independent groups reported an X-ray excess at an energy of $E_{\gamma}=3.55$ keV using \emph{Chandra} and \emph{XMM-Newton} data from galaxy clusters and Andromeda \cite{Bulbul:2014sua,Boyarsky:2014jta}. The excess, being consistent with a line feature, was speculated to originate from the decay of a $m_{\rm sn}=7.1$ keV sterile neutrino. Both the presence of the suggested signal and it's interpretation are disputed. First of all, the excess has not been found in dwarf galaxy observations \cite{Malyshev:2014xqa,Anderson:2014tza,Jeltema:2015mee,Ruchayskiy:2015onc}. Furthermore, it was argued that the signal could arise due to an incomplete subtraction of atomic lines \cite{Jeltema:2014qfa,Phillips:2015wla} and that it does not have the right morphology expected by the DM profile \cite{Urban:2014yda,Carlson:2014lla} (although both criticisms have been challenged as well \cite{Boyarsky:2014ska,Iakubovskyi:2015dna}).

Currently, it seems very hard to draw definite conclusions from the status of the emission line, as poor statistics leave room for interpretation. However, upcoming X-ray surveys will improve the situation and could potentially clarify the origin of the emission line \cite{Iakubovskyi:2014yxa,Neronov:2015kca}.

\subsection{Constraints}
There is a large number of studies reporting limits on the sterile neutrino parameter space based on X-ray data. Examples are \emph{Chandra} observations of Andromeda \cite{Watson:2011dw}, \emph{XMM-Newton} observations of the LMC \cite{Boyarsky:2006fg} and Willman 1 \cite{Loewenstein:2012px}, or \emph{Suzaku} observations of the Perseus cluster and the diffuse background \cite{Tamura:2014mta,Sekiya:2015jsa}. More detailed summaries of X-ray limits (prior to 2013) can be found in \cite{Canetti:2012kh,Merle:2013ibc}.

In this paper we use recent bounds from \emph{Suzaku} data of the diffuse background \cite{Sekiya:2015jsa} and the Perseus cluster \cite{Tamura:2014mta} (thick solid black line in Figs.~\ref{fig:xmean} and~\ref{fig:constraints}), the former providing the limit below and the latter above $m_{\rm sn}\sim10$ keV. The  \emph{Suzaku} constraint from the diffuse background is similar in strength to the {\it Chandra} limits from Andromeda \cite{Watson:2011dw} and somewhat weaker than recent constraints from \cite{Horiuchi:2013noa}, \cite{Riemer-Sorensen:2014yda}, and \cite{Malyshev:2014xqa} based on observations of Andromeda, the Milky-Way centre, and a collection of nearby dwarf galaxies. We therefore conclude that it is a rather conservative estimate which can be safely adopted as a reference line for this work.

The {\it Suzaku} bound from the Perseus cluster, on the other hand, is significantly stronger than former limits at mass scales beyond 10 keV. We therefore provide an additional, more conservative limit at the high-mass end (dashed black line in Figs.~\ref{fig:xmean} and~\ref{fig:constraints}) which is based on a compilation of results from Refs.~\cite{Canetti:2012kh} and \cite{Ng:2015gfa}.

%%%%%%%%%%%%%%%%%%%%%%%%%
\section{\label{sec:structform}Structure formation}
Sterile neutrinos are a non-cold DM candidate and therefore likely to modify structure formation at the smallest observable scales. For this reason they have been suggested as a solution to possible small-scale issues of $\Lambda$CDM, such as the {\it missing satellite} \cite{Bode:2000gq,Dolgov:2000ew,Zavala:2009ms,Reed:2014cta} or the {\it too-big-to-fail} problem \cite{Lovell:2011rd,Anderhalden:2012jc,Schneider:2013wwa}. On the other hand, the reduced small-scale clustering of sterile neutrino DM can be used to constrain the sterile neutrino mass. The most stringent constraints come from the analysis of the Lyman-$\alpha$ forest, the characteristic absorption lines from distant quasars. Independent limits are obtained via the abundance of dwarf galaxies in the local neighbourhood. However, most of the constraints published so far refer to warm DM (WDM) with a characteristic power suppression from a thermal-like (Fermi-Dirac) momentum distribution, while sterile neutrinos may have shallower suppressions. In this section we summarise constraints from both Lyman-$\alpha$ and Milky-Way satellites, and we show how they can be applied to more general power suppressions.

\subsection{\label{sec:lyal}Constraints from the Lyman-$\mathbf{\alpha}$ forest}
Lyman-$\alpha$ absorption lines probe the distribution of neutral hydrogen along the line-of-sight of luminous, high-redshift quasars. Since the hydrogen is a tracer of the underlying DM perturbations, Lyman-$\alpha$ observations can be used to test cosmology.

The most convenient statistics from the Lyman-$\alpha$ forest is the flux power spectrum $P_{F}(k_v)$ which is related to the one-dimensional (1D) matter power spectrum $P_{\rm 1D}(k)$ by the bias function $b^2(k)=P_{F}(k_v)/P_{\rm 1D}(k)$, where $k=aHk_v$ ($H$ being the Hubble parameter for a sale-factor $a$). Assuming an isotropic matter distribution \cite{Zaldarriaga:2001xs}, the 1D power spectrum is given by \cite{Kaiser:1990xe}
\begin{equation}\label{PS1d}
P_{\rm 1D}(k)=\frac{1}{2\pi}\int_{k}^{\infty}dk' k' P(k'),
\end{equation}
where $P(k)$ is the three-dimensional (3D) matter power spectrum. The flux power spectrum from high-redshift quasars typically spans $k$-ranges of $0.1-2$ h/Mpc,  but much smaller scales (down to the Jeans length of the gas) are probed via the integral of Eq.~\ref{PS1d} (which formally extends to arbitrarily high $k$-values). This further increases the sensitivity of the Lyman-$\alpha$ forest as a probe for small-scale structure formation.

Over the last fifteen years several studies have put forward constraints on non-resonant sterile neutrino DM (or equivalently thermal relic WDM) based on Lyman-$\alpha$ data. An early study by Ref.~\cite{Narayanan:2000tp} ruled out thermal relic masses below $m_{\rm WDM}\sim0.75$ keV. About five years later Refs.~\cite{Seljak:2006qw,Viel:2006kd} found more stringent constraints of $m_{\rm WDM}\gtrsim2.5$ keV based on SDSS data, which allowed to rule out non-resonantly (DW) produced sterile neutrinos as DM candidate.

Recently, Ref.~\cite{Viel:2013apy} (hereafter V13) used high-redshift quasar spectra from MIKE and HIRES and reported a 2-$\sigma$ bound of $m_{\rm WDM}\gtrsim3.3$ keV on thermal relic WDM (corresponding to a non-resonant sterile neutrino mass of $m_{\rm sn}\gtrsim18.5$ keV, see \cite{Bozek:2015bdo}). This improvement was possible despite comparatively poor statistics (i.e a total of 25 spectra) owing to unprecedented data range in scale and redshift. Even tighter constraints were obtained by Ref.~\cite{Baur:2015jsy} (hereafter B15), who used more than 13000 quasar spectra from the BOSS survey. B15 found a 2-$\sigma$ limit of $m_{\rm WDM}\gtrsim4.35$ keV (i.e. a non-resonant sterile neutrino mass limit of $m_{\rm sn}\gtrsim26.4$ keV\footnote{B15 quote a limit of $m_{\rm sn}\gtrsim31.7$ keV based on the mass conversion from Ref.~\cite{Viel:2005qj}. Here we give a lower value in agreement with the updated conversion from Ref.~\cite{Bozek:2015bdo}}) which is the most stringent limit on thermal relic WDM up to date.

While DW sterile neutrinos have been ruled out via the combination of Lyman-$\alpha$ and X-ray data, there are no strong constraints on resonantly produced sterile neutrinos so far. This is mainly due to the nontrivial shape of the power suppression, which requires a detailed investigation of the parameter space. The authors of Ref. \cite{Boyarsky:2008xj} have derived bounds on a mixed DM scenario (i.e a combination of CDM and thermal relic WDM) arguing that such a model provides a good estimate for resonant sterile neutrino DM. They showed that the spectra of the two scenarios are indeed very similar around the scales where the suppression becomes visible, but they deviate substantially at even smaller scales. The constraints from~\cite{Boyarsky:2008xj} are therefore expected to be weaker than what can be obtained by directly studying resonant sterile neutrino DM.

In this section we present a new method to constrain resonant sterile neutrino DM using limits obtained with non-resonant models. While absolute bounds can only be computed with a full statistical analysis (which is beyond the scope of this paper), it is nevertheless possible to determine if a given resonant scenario deviates more or less from the CDM baseline than a non-resonant reference model. If the deviation is consistently larger (over the entire range probed by Lyman-$\alpha$), then the scenario is excluded at a higher confidence level (CL) as the reference model. 

For the non-resonant reference we choose the 2-$\sigma$ excluded models from V13 and B15 characterised by thermal relic masses of $m_{\rm WDM}=3.3$ keV and $m_{\rm WDM}=4.35$ keV, respectively. This allows us to ultimately quote 2-$\sigma$ limits for the full parameter space of resonantly produced sterile neutrino DM.

We now give a detailed prescription of the method sketched above: (i) We compute the 3D linear matter power spectra of resonant sterile neutrino DM models (distributed on a grid over the parameter space) using the Boltzmann solver {\tt CLASS} \cite{Lesgourgues:2011re,Blas:2011rf,Lesgourgues:2011rh} with {\it Planck} values for the cosmological parameters \cite{Planck:2015xua}. There is no need to run simulations, since we are not interested in absolute power spectra but only in the relative hierarchy of different scenarios. (ii) We use Eq.~(\ref{PS1d}) to calculate 1D power spectra over the entire range probed by the Lyman-$\alpha$ observations. This depends on the corresponding data and is $0.5 \,{\rm h/Mpc}\lesssim k\lesssim 10 \,{\rm h/Mpc}$ for the spectra from MIKE+HIRES used in V13 and $0.2 \,{\rm h/Mpc}\lesssim k\lesssim 2 \,{\rm h/Mpc}$ for the BOSS spectra \cite{Palanque-Delabrouille:2015pga} used in B15. (iii) We define the ratio
\begin{equation}\label{ratio}
r(k)=\frac{P_{\rm 1D}(k)}{P_{\rm 1D}^{\rm REF}(k)},
\end{equation}
where $P_{\rm 1D}(k)$ is the 1D power spectrum of a given resonant scenario, while $P_{\rm 1D}^{\rm REF}(k)$ corresponds to the 1D power spectrum of the non-resonant reference model from either V13 or B15. (iv) We use the ratio $r(k)$ to determine whether a given resonant scenario is allowed or excluded. This is only possible because the bias factor (defined above Eq.~\ref{PS1d}) differs very little between CDM and WDM models \cite{Viel:2005qj}, which means that Eq.~(\ref{ratio}) describes the ratio of flux power spectra as well. We define a model to be excluded (at 2-$\sigma$ CL), if the ratio is both below one and monotonically decreasing over the entire range probed by Lyman-$\alpha$. The two requirements guarantee that the absolute {\it value} and the {\it slope} of the given resonant scenario deviates more from the CDM model than the non-resonant reference model.

The reason why we use both {\it value} and {\it slope} of the ratio $r(k)$ as requirements, is that the simulated mean Lyman-$\alpha$ flux is forced to the observed value for all models in the analysis of V13 and B15. This results in a renormalisation of the 1D power spectrum with slightly more power for WDM with respect to CDM at large scales (see e.g. \cite{Viel:2013apy} for a more detailed explanation). By considering the {\it slope} as second requirement, we guarantee to discard models independently of this normalisation.

A similar approach to rule out models with non-trivial power spectra has been developed by the authors of Ref.~\cite{Palazzo:2007gz}. They used Lyman-$\alpha$ constraints from non-resonant sterile neutrino DM (obtained by Ref.~\cite{Seljak:2006qw}) to evaluate the validity of mixed DM scenarios (i.e a mixture of warm and cold DM). Their method consists of defining a pivot scale $k_p=2.0$ h/Mpc  (which corresponds to the smallest scale probed by the Lyman-$\alpha$ spectra from SDSS) and discarding all scenarios with less 1D power than the reference model (i.e. the 2-$\sigma$ excluded model from Ref.~\cite{Seljak:2006qw}) at that pivot scale. This approach is similar but less stringent than our method, as it only considers one scale ($k_p$) and does not include the spectral slope.

\begin{figure}[!ht]
\center{
\includegraphics[width=.325\textwidth,trim={1.2cm 1.1cm 3.4cm 0cm}]{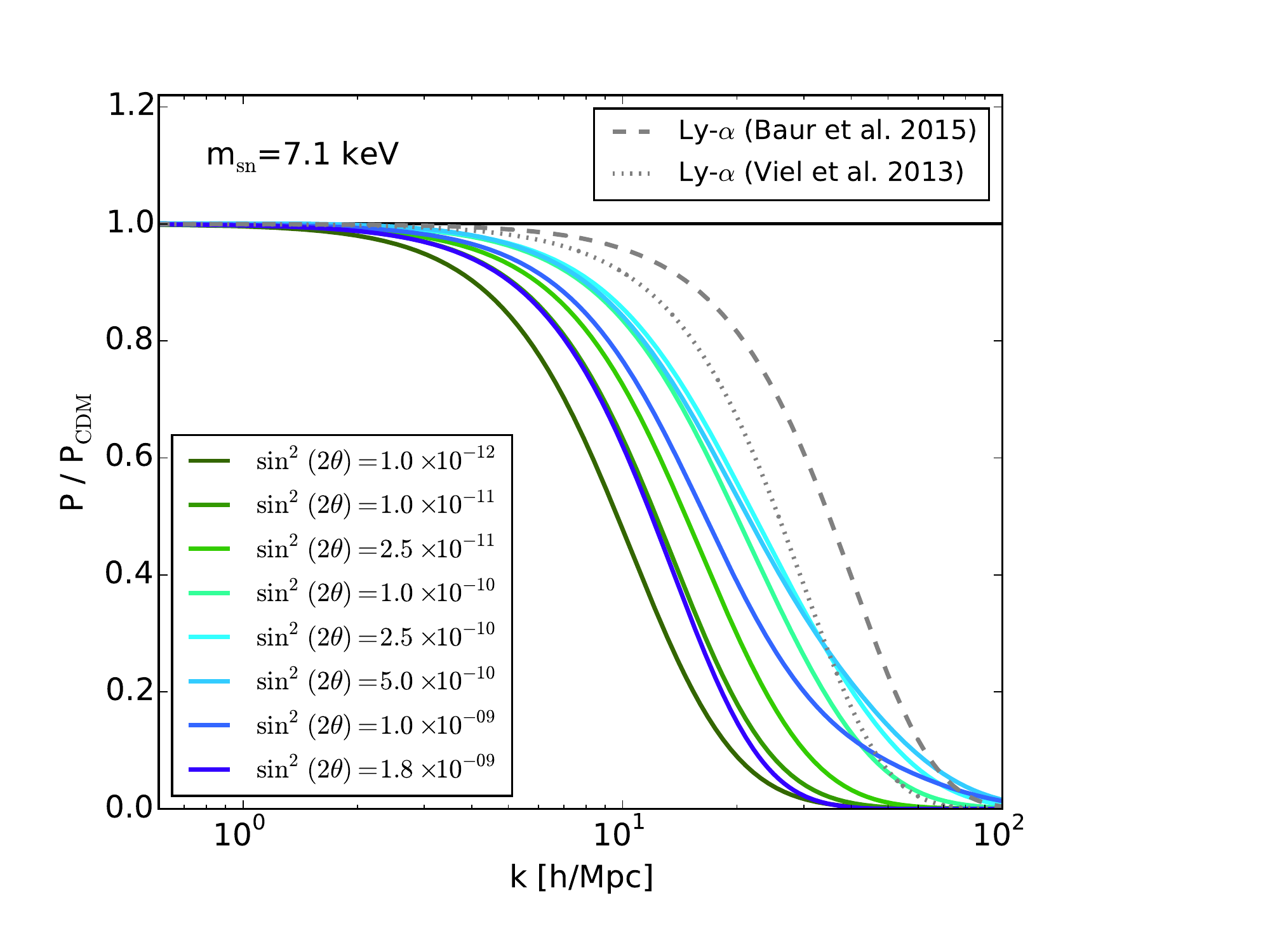}
\includegraphics[width=.325\textwidth,trim={1.2cm 1.1cm 3.4cm 0cm}]{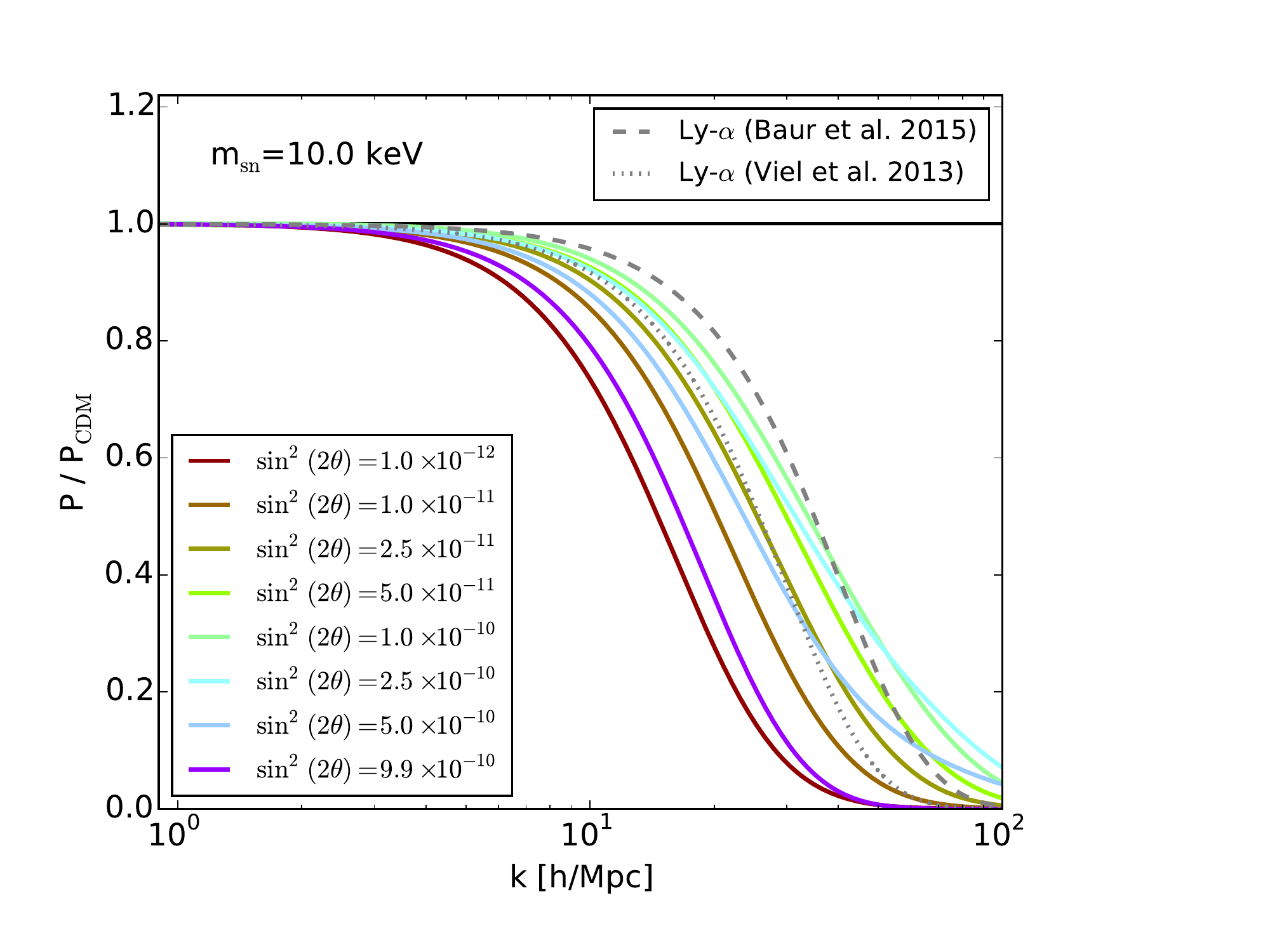}
\includegraphics[width=.325\textwidth,trim={1.2cm 1.1cm 3.4cm 0cm}]{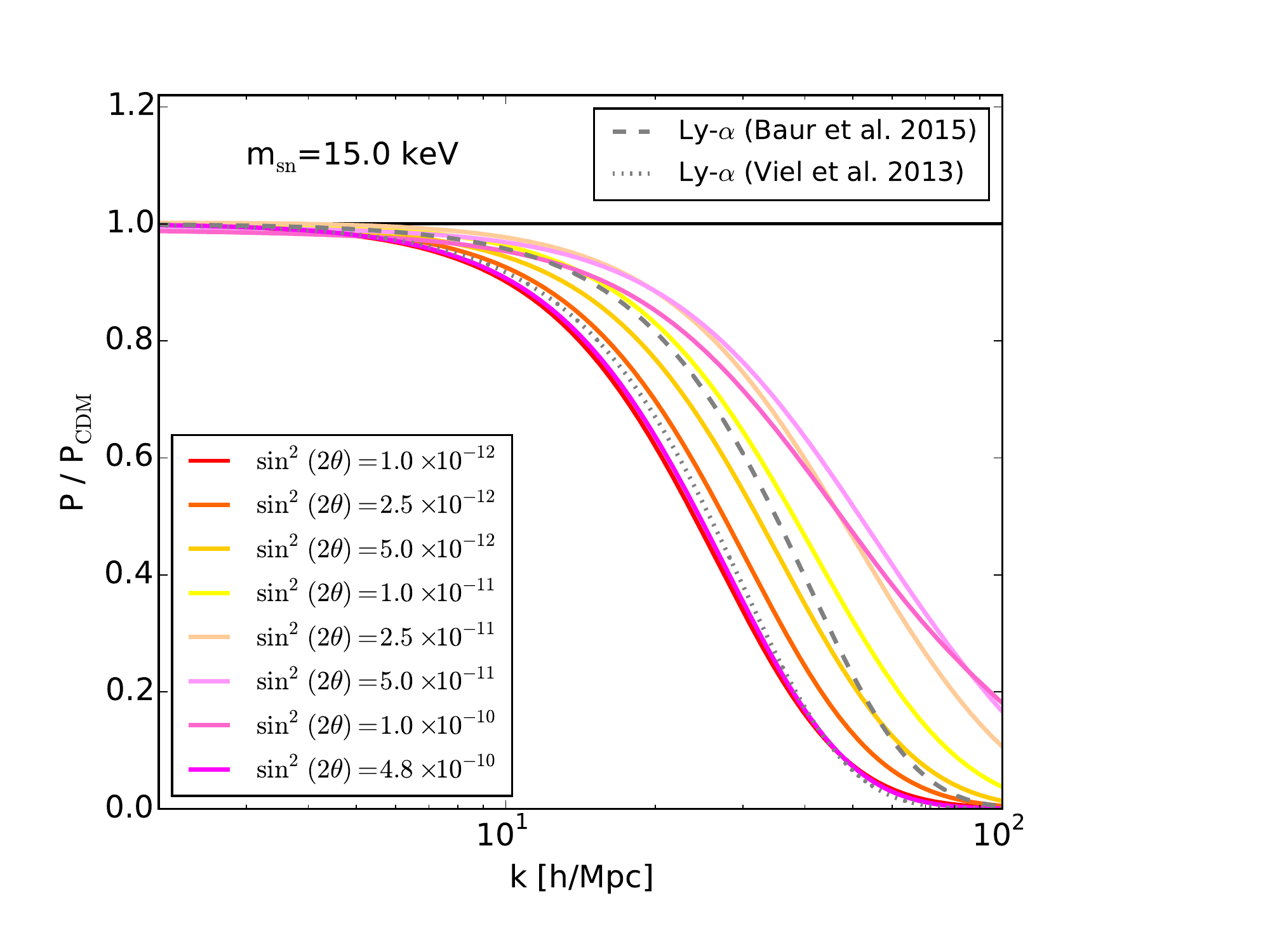}\\
\includegraphics[width=.325\textwidth,trim={1.2cm 1.1cm 3.4cm 0cm}]{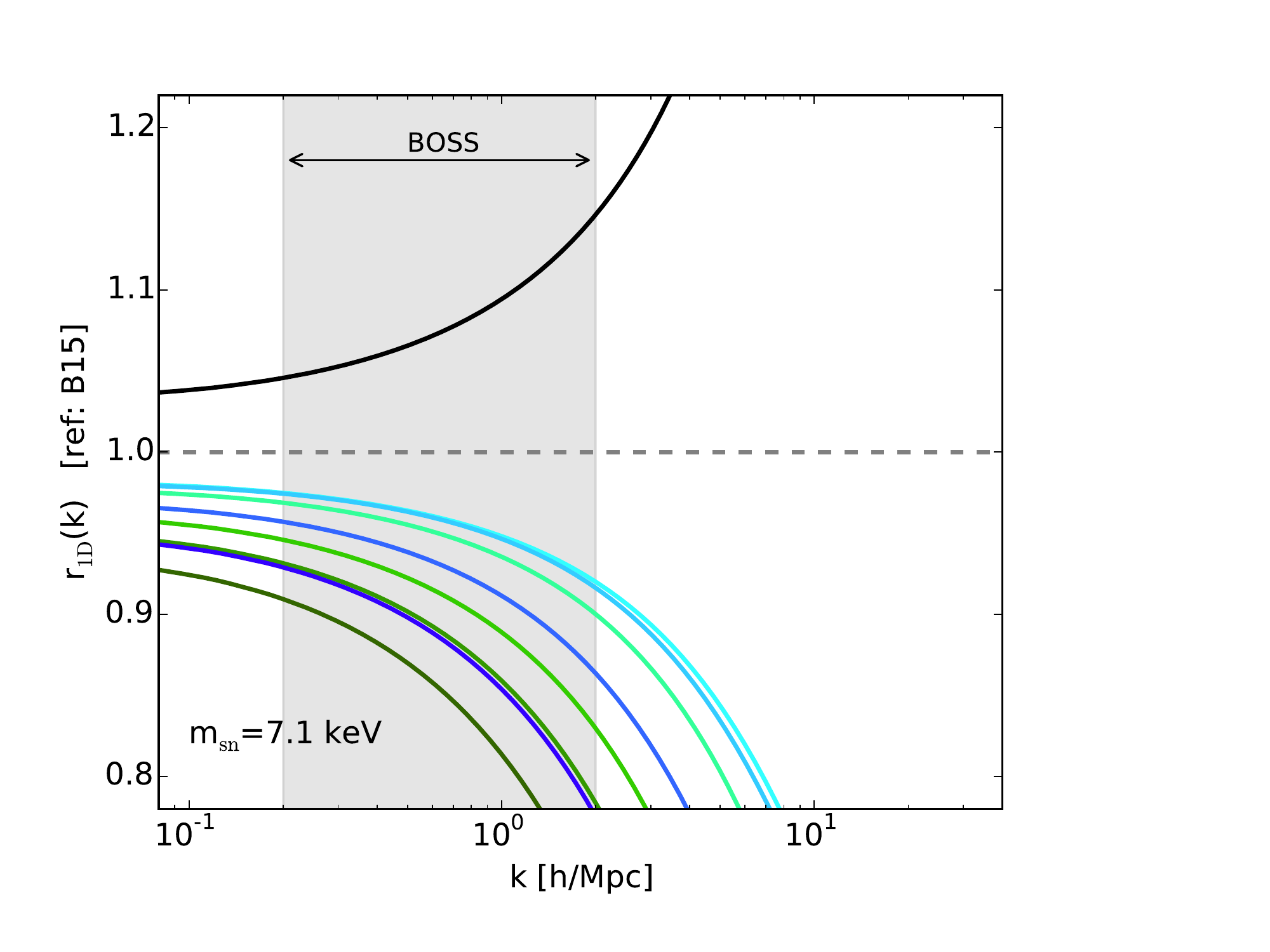}
\includegraphics[width=.325\textwidth,trim={1.2cm 1.1cm 3.4cm 0cm}]{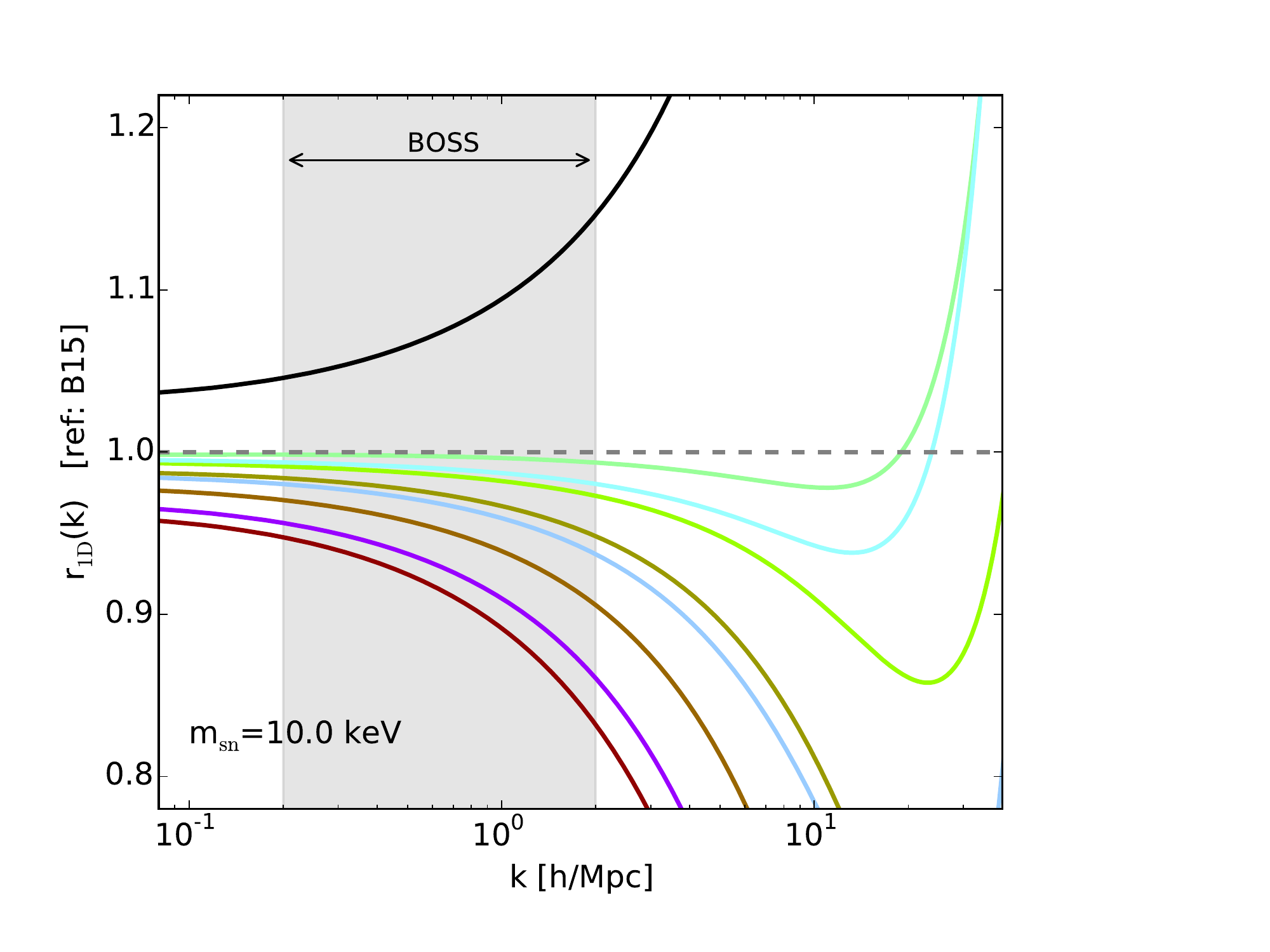}
\includegraphics[width=.325\textwidth,trim={1.2cm 1.1cm 3.4cm 0cm}]{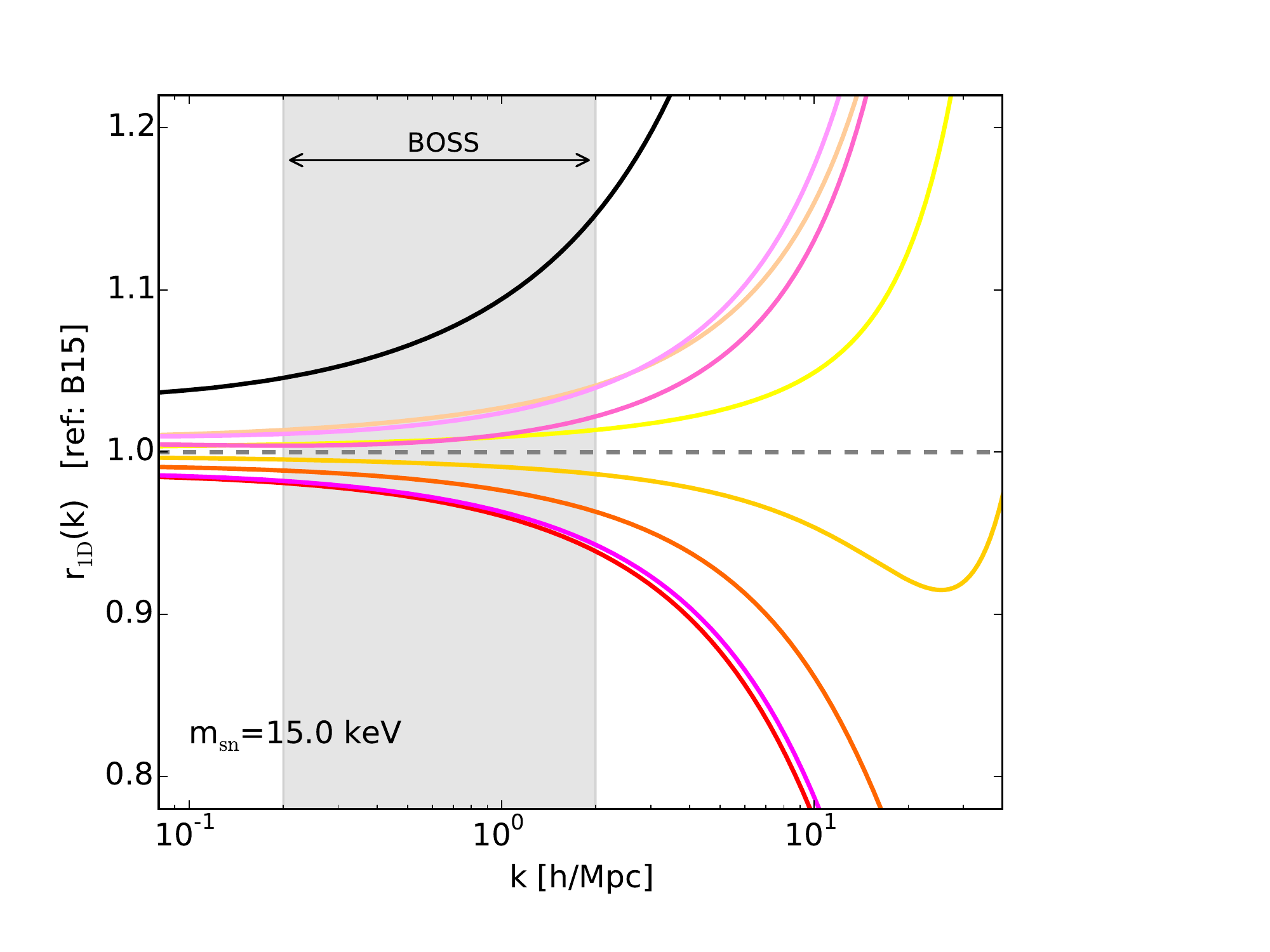}\\
\includegraphics[width=.325\textwidth,trim={1.2cm 1.1cm 3.4cm 0cm}]{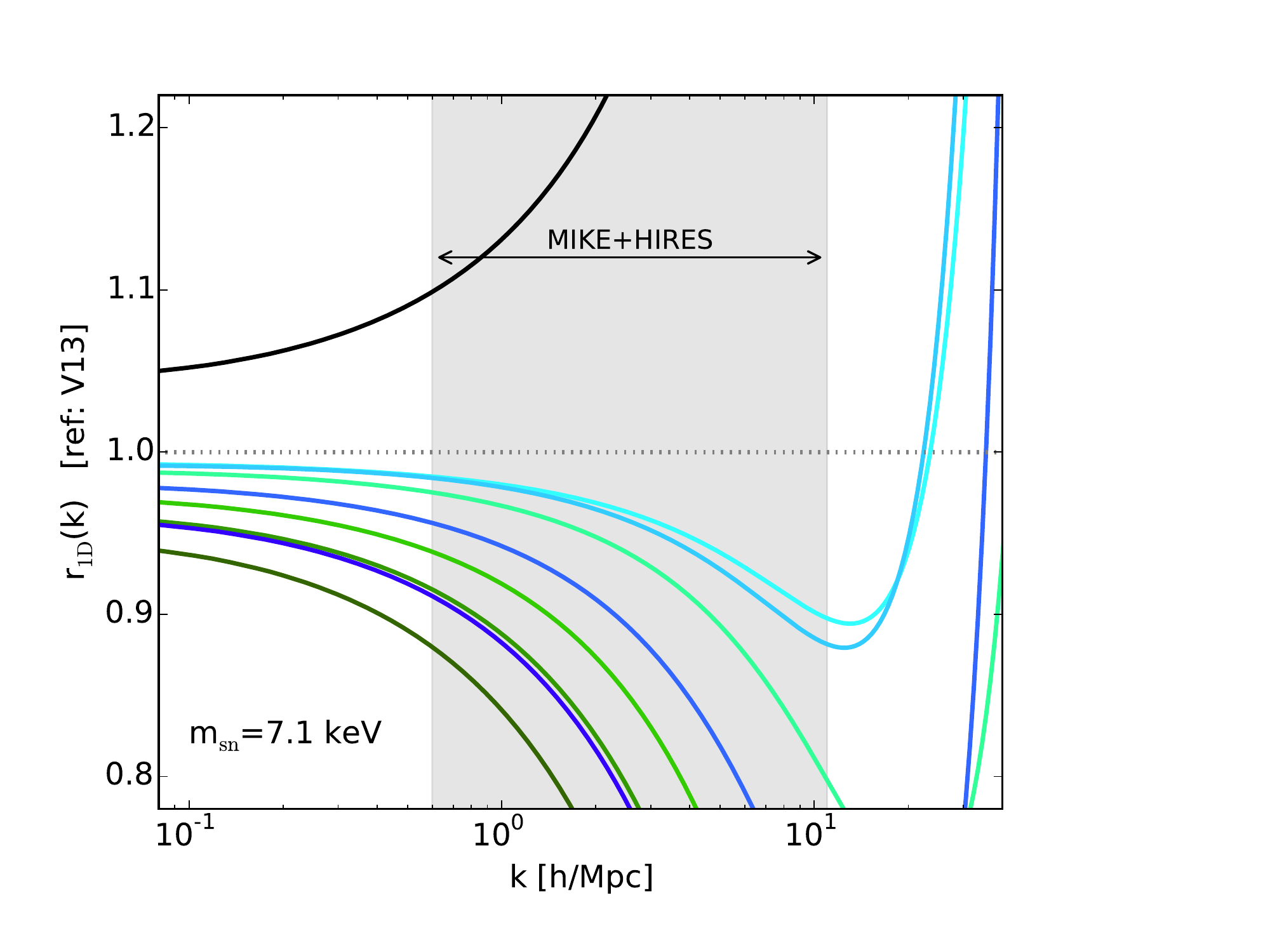}
\includegraphics[width=.325\textwidth,trim={1.2cm 1.1cm 3.4cm 0cm}]{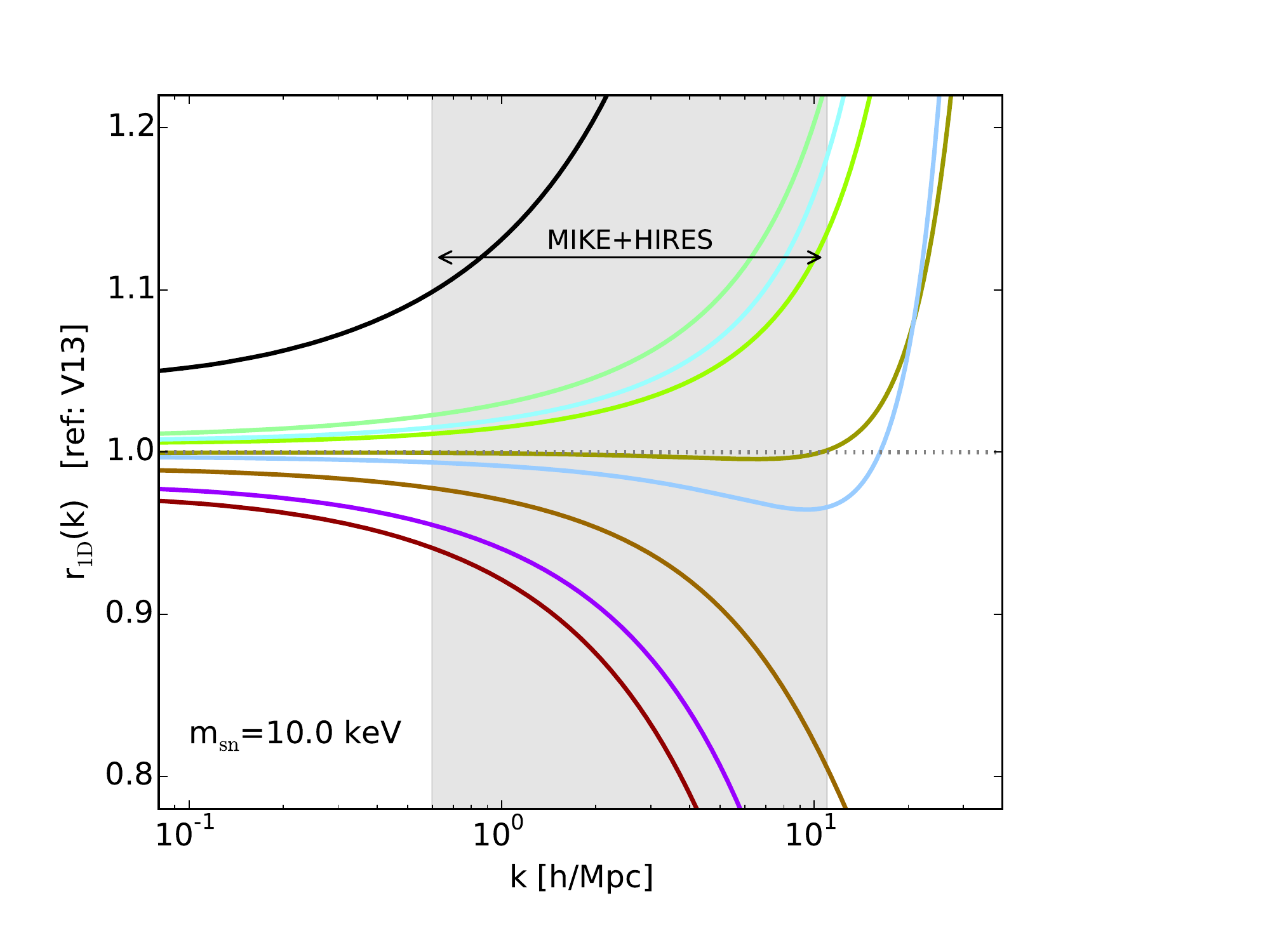}
\includegraphics[width=.325\textwidth,trim={1.2cm 1.1cm 3.4cm 0cm}]{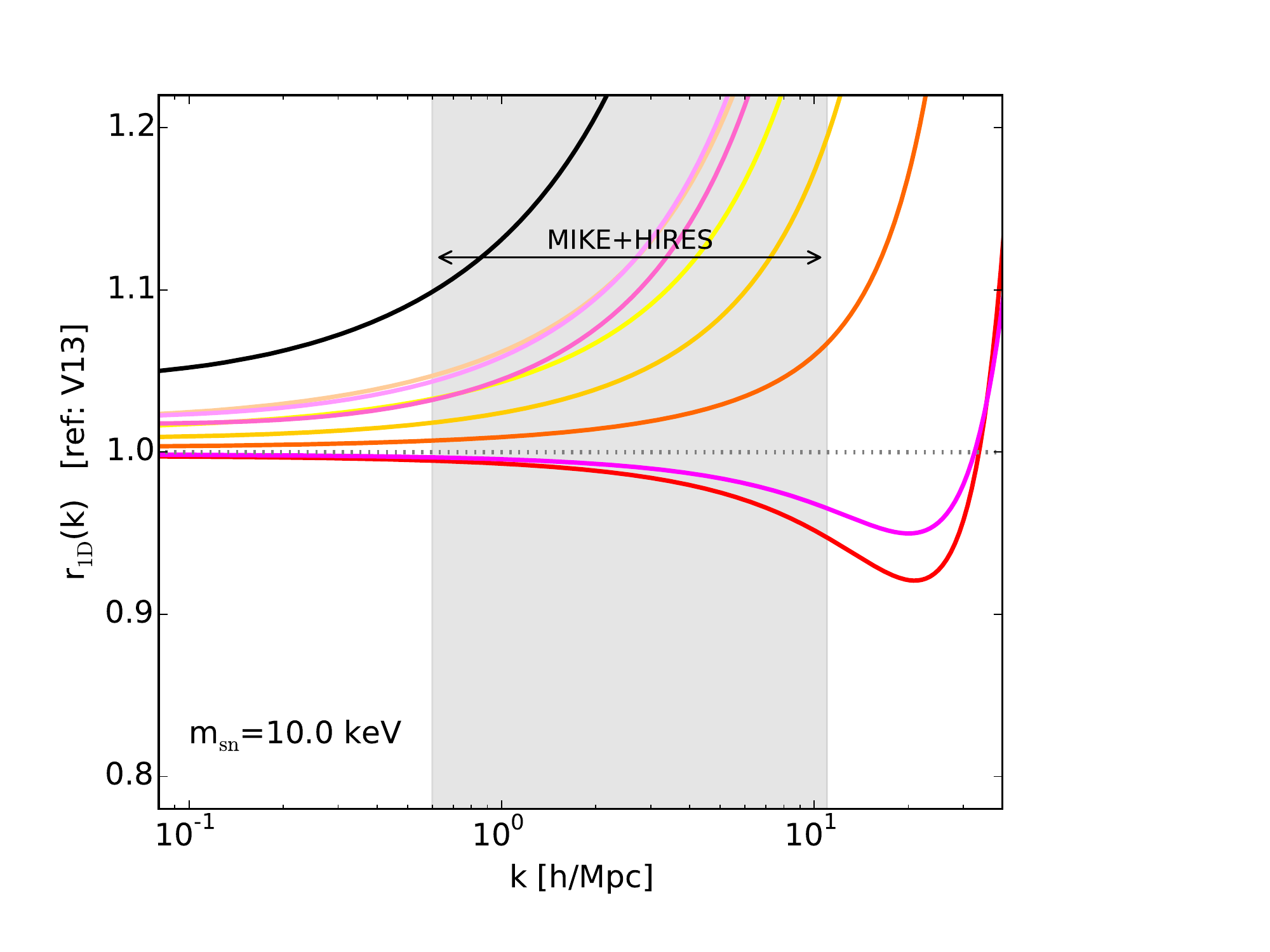}
\caption{\label{fig:ps}Relative power spectra of the resonant models from Fig.~\ref{fig:dis} (same colour scheme). {\it Top}: 3D matter power spectra with respect to the CDM baseline. The non-resonant reference models of B15 and V13 (excluded at 2-$\sigma$ CL) are shown as dashed and dotted grey lines. {\it Middle:}  Ratios of the 1D power spectra with respect to the B15 reference model. {\it Bottom}: Same for the V13 reference model. Resonant scenarios are excluded (at 2-$\sigma$ CL) if they lie below one and if they decrease monotonically over the entire range covered by the Lyman-$\alpha$ data (grey areas representing BOSS data for B15 and MIKE+HIRES data for V13).}}
\end{figure}

In Fig.~\ref{fig:ps} we plot the power spectra of resonant DM models with varying mass and mixing angles. They are the same cases than those presented in Fig.~\ref{fig:dis}. The top panels show ratios of the 3D matter power spectra with respect to the CDM baseline. The reference models of B15 and V13 (corresponding to a thermal relic WDM mass of 4.35 keV and 3.3 keV, respectively) are shown as dashed and dotted grey lines. Based on these plots it is already possible to rule out several scenarios which have less power than the reference models over all scales. However, there are many scenarios with less power at smaller and more power at larger $k$-values, owing to the fact that resonant production tends to yield shallower power suppressions. These cases require the more thorough analysis described above.

The middle panels of Fig.~\ref{fig:ps} show ratios of the 1D power spectra (given by Eq.~\ref{ratio}) with respect to the B15 reference model (i.e. thermal relic WDM with $m_{\rm WDM}=4.35$ keV). The shaded area designates the range of scales of the BOSS Lyman-$\alpha$ data used by B15. Models are excluded if they exhibit decreasing ratios below one within the grey area. This is the case for all models except the ones with the coldest momenta at $m_{\rm sn}=15$ keV.

The bottom panels of Fig.~\ref{fig:ps} show the same scenarios, now with respect to the reference model of V13 (i.e. thermal relic WDM with $m_{\rm WDM}=3.3$ keV). The shaded area now corresponds to the Lyman-$\alpha$ data from MIKE+HIRES. As expected, the V13 limits are less restrictive, excluding all models with $m_{\rm sn}=7.1$ keV but only some at larger particle mass.

It is worth noting that all models formally excluded by the recipe explained above have 3D power spectra with visually stronger suppressions (starting at lower $k$-values) than the corresponding non-resonant reference models. This is not surprising but consists of an important cross-check confirming the validity of our approach.

%%%%%%%%%%%%%%%%%%%%%%%%%
\subsection{\label{sec:dwarfs}Constraints from Milky-Way satellite counts}
Dwarf galaxy counts around the Milky Way provide a powerful test for alternative dark matter scenarios. It is well known that in CDM the sub-halo abundance largely exceeds the number of observed satellites within the Milky-Way. This discrepancy can be quite easily remedied by assuming the right amount of photo-heating during the reionization period to push gas out of small haloes and shut down star formation. As a consequence, in the CDM scenario one expects a large number of unobserved dark subhaloes to orbit within the Milky-Way potential.

Despite rather large uncertainties due to poorly understood baryonic effects, it is nevertheless possible to use observed satellite numbers as a test for non-cold DM scenarios. While there could be many \emph{more} unobservable sub-haloes orbiting the Milky-Way, it is not possible to have \emph{less} sub-haloes than satellites observed in the sky. This puts a stringent limit on the coldness of the DM fluid and has been used to constrain thermal relic WDM in the past (see for example~\cite{Polisensky:2010rw,Kennedy:2013uta,Schneider:2014rda}).

There are several potential systematic errors that need to be accounted for when using satellites to constrain DM models. First of all, the number of sub-haloes depends on the mass of the Milky-Way \cite{Kennedy:2013uta}, which is not very well known \cite{Guo:2009fn}. Furthermore, observed satellites need to be assigned to predicted subhaloes, either by estimating the satellite mass (or circular velocity) \cite{Polisensky:2010rw} or by modelling star formation and comparing luminosities \cite{Kennedy:2013uta}. Finally, there is an object-to-object scatter of satellite numbers per host that needs to be taken into account.

In this paper we follow the approach presented in Ref.~\cite{Schneider:2014rda} and us an extended Press-Schechter approach (based on the sharp-$k$ filter, see \cite{Schneider:2013ria,Benson:2012su,Schneider:2014rda}) to estimate the number of sub-haloes for a given DM scenario. The relation
\begin{equation}\label{Nsub}
\frac{{\rm d}N_{\rm sh}}{{\rm d} \ln M_{\rm sh}}=\frac{1}{C_{\rm n}}\frac{1}{6\pi^2} \left(\frac{M_{\rm hh}}{M_{\rm sh}}\right)\frac{P(1/R_{\rm sh})}{R_{\rm sh}^3\sqrt{2\pi(S_{\rm sh}-S_{\rm hh})}} \,,
\end{equation}
is based on the conditional mass function normalised to $N$-body results (see \cite{Schneider:2014rda} for more details). The variance $S_i$ and mass $M_i$ are defined as
\begin{equation}
S_{i}=\frac{1}{2\pi^2}\int_0^{1/R_i}{\rm d}k\ k^2P(k),\hspace{0.5cm}M_i= \frac{4\pi}{3}\Omega_{\rm m} \rho_c(cR_i)^3,\hspace{0.5cm}c=2.5,\hspace{1cm}i=\{\rm sh,hh\},
\end{equation}
where sh and hh stand for {\it subhalo} and {\it host-halo}, respectively. The normalisation constant $C_{\rm n}$ depends on the halo definition, being $C_{\rm n}=45$ if the host halo is delimited by a density threshold of 200 times the background density and $C_{\rm n}=34$ for a threshold of 200 times the critical density. Eq. \eqref{Nsub} solely depends on the linear power spectrum $P(k)$ for a given DM scenario and provides a good prescription of the predicted sub-halo abundance.

We estimate the number of satellites within the Milky-Way following the procedure of \cite{Polisensky:2010rw} (but see also \cite{Kennedy:2013uta,Schneider:2014rda,Merle:2015vzu}). This consists of adding up the 11 classical satellites (assumed to be a complete sample at their mass range) with the 15 ultra-faint satellites from SDSS which are multiplied by a factor of 3.5 to account for the limited sky coverage of the SDSS survey. This results in an estimated satellite count of $63$ within the virial radius of the Milky-Way. In order to account for the halo-to-halo scatter we further reduce this number by ten percent (see e.g. Ref.~\cite{Mao:2015yua}) ending up with the conservative estimate of $N_{\rm Sat}>57$. More recently, DES has found several fainter satellite candidates below the detection limit of SDSS \cite{Koposov:2015cua,Bechtol:2015cbp,Drlica-Wagner:2015ufc}, suggesting the presence of many more undetected satellites even in the SDSS field-of-view. However, we ignore these new findings, as there are no reliable mass estimates up to date and some of the candidates could be globular clusters instead.

The expected number of satellites is proportional to the Milky-Way mass (see Eq.~\ref{Nsub}), which is only known within a factor of a few. For this study we take an upper limit of $M_{\rm hh}<3\times 10^{12}$ M$_{\odot}$/h defined with respect to 200 times the background density. This is a conservative limit based on various direct mass estimates from the literature (see Ref.~\cite{Wang:2015ala} for a summary). The total mass of the observed satellites can in principle be estimated with the help of stellar kinematics. Due to rather large uncertainties, we only assume all dwarfs to have total masses above $M_{\rm sh}=10^8$ M$_{\odot}$/h (which again is a conservative estimate, see Ref.~\cite{Brooks:2012vi}). 

\begin{table}
\ra{1.2}
   \centering{
\begin{tabular}{cr||cr||cr}
  %\hline
  $m_{\rm sn}= 7.1$ keV & & $m_{\rm sn}=10$ keV & & $m_{\rm sn}=15$ keV & \\
  \hline\hline
  $\sin^22\theta$ & $N_{\rm sh}^{\rm tot}$ & $\sin^22\theta$ & $N_{\rm sh}^{\rm tot}$ &$\sin^22\theta$ & $N_{\rm sh}^{\rm tot}$ \\
  \hline
  $\mathbf{1.0\times10^{-12}}$ & $\mathbf{18}$ &$\mathbf{1.0\times10^{-12}}$ & $\mathbf{44}$ &$5.0\times10^{-13}$ & $102$ \\
  $\mathbf{1.0\times10^{-11}}$ & $\mathbf{30}$ &$1.0\times10^{-11}$ & $91$ &$1.0\times10^{-12}$ & $112$ \\
  $\mathbf{2.5\times10^{-11}}$ & $\mathbf{50}$ &$2.5\times10^{-11}$ & $127$ &$2.5\times10^{-12}$ & $137$ \\
  $1.0\times10^{-10}$ & $94$ &$5.0\times10^{-11}$ & $152$ &$5.0\times10^{-12}$ & $162$ \\
  $2.5\times10^{-10}$ & $114$ &$1.0\times10^{-10}$ & $168$ &$1.0\times10^{-11}$ & $185$ \\
  $5.0\times10^{-10}$ & $112$ &$2.5\times10^{-10}$ & $160$ &$5.0\times10^{-11}$ & $212$ \\
  $1.0\times10^{-09}$ & $78$ &$5.0\times10^{-10}$ & $121$ &$1.0\times10^{-10}$ & $202$ \\
  $\mathbf{1.8\times10^{-09}}$ & $\mathbf{24}$ &$\mathbf{9.9\times10^{-10}}$ & $\mathbf{56}$ &$4.8\times10^{-10}$ & $117$ \\
\end{tabular}}
\caption{Number of sub-haloes (with mass above $10^{8}$ M$_{\odot}$/h) orbiting a host halo of $M_{\rm hh}=3\times10^{12}$ M$_{\odot}$/h (i.e., the upper limit for the Milky-Way mass, see text) for the scenarios shown in Figs.~\ref{fig:dis} and~\ref{fig:ps}. Excluded scenarios (with $N_{\rm sh}^{\rm tot}<57$) are highlighted in bold. For CDM we obtain $N_{\rm sh}^{\rm tot}=265$.}\label{satnum}
\end{table}

The predicted total number of sub-haloes above $10^8$ M$_{\odot}$/h ($N_{\rm sh}^{\rm tot}$) is obtained by integrating over Eq.~(\ref{Nsub}). In Table \ref{satnum} we list $N_{\rm sh}^{\rm tot}$ for all scenarios illustrated in Fig.~\ref{fig:dis} and \ref{fig:ps} assuming $M_{\rm hh}=3\times10^{12}$ M$_{\odot}$/h (i.e the upper limit for the Milky-Way mass). Only some of the 7.1 keV and 10 keV but none of the 15 keV scenarios can be excluded (i.e. obey $N_{\rm sh}^{\rm tot}<N_{\rm Sat}$). This shows that the constraints from Milky-Way satellite counts are somewhat less stringent than the ones from the Lyman-$\alpha$ forest.

%%%%%%%%%%%%%%%%%%%%%%%%%%%%%

\begin{figure}[!ht]
\center{
\includegraphics[width=.99\textwidth,trim={1.1cm 0cm 2.0cm 0cm}]{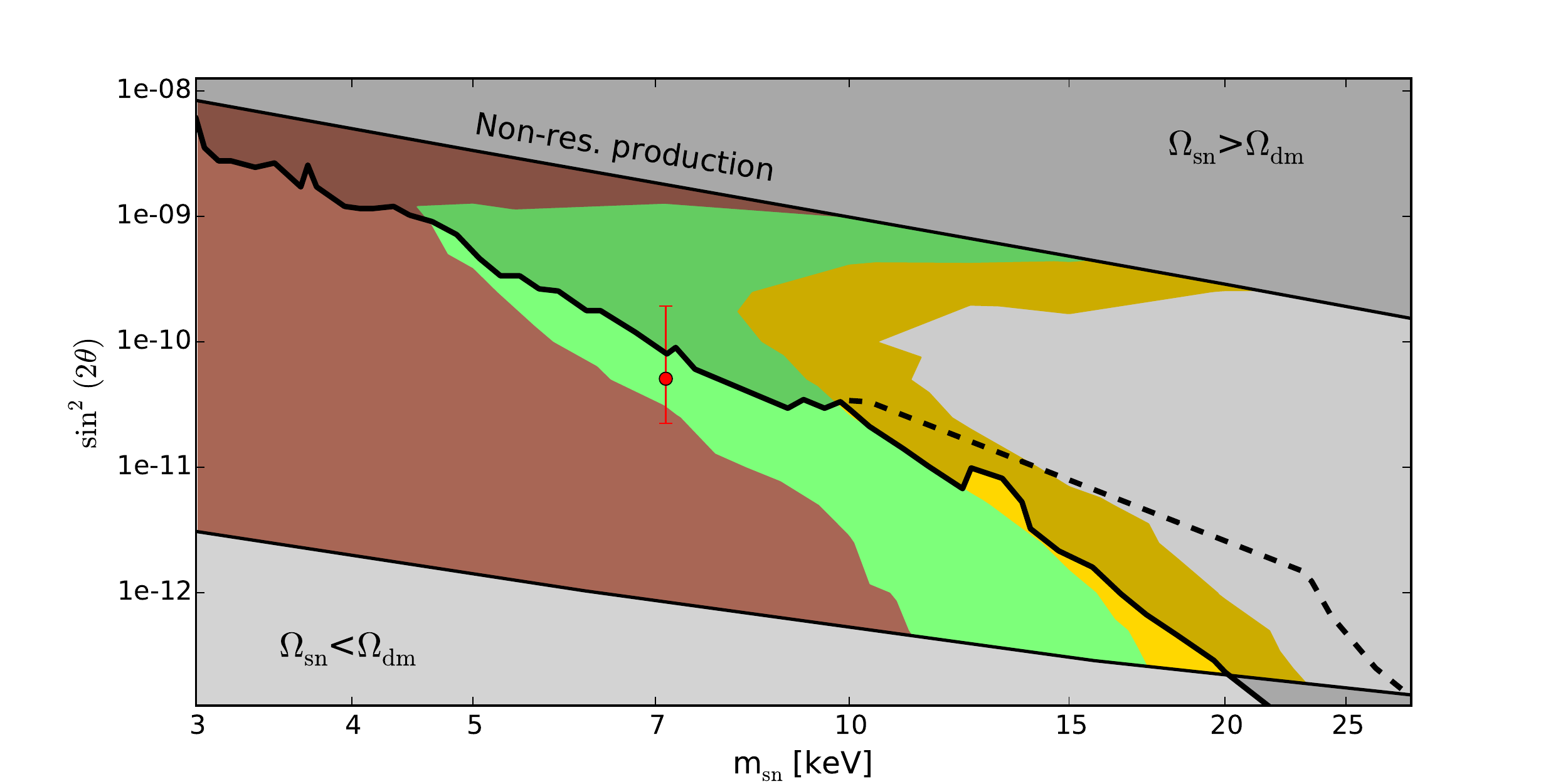}
\caption{\label{fig:constraints}Constraints from structure formation on the sterile neutrino parameter space. The areas in green and yellow are excluded by Lyman-$\alpha$ bounds (based on the V13 and B15 reference models, see Sec.~\ref{sec:lyal}). The brown area is excluded by Milky-Way satellite counts (see Sec.~\ref{sec:dwarfs}). The parameter space is delimited by an upper and lower thin line corresponding to zero (non resonant production) and maximum lepton asymmetry. The thick line illustrates the X-ray constraints from {\it Suzaku} \cite{Tamura:2014mta,Sekiya:2015jsa}, the dashed line an independent X-ray limit from Refs.~\cite{Canetti:2012kh,Ng:2015gfa}.  The tentative line signal~\cite{Bulbul:2014sua,Boyarsky:2014jta} at 7.1 keV is shown by the red symbol.}}
\end{figure}

\section{\label{sec:results}Results and Discussion}
In the last section we have discussed how observations of the Lyman-$\alpha$ forest and Milky-Way satellites can be used to constrain resonantly produced sterile neutrino DM. We now apply these methods to the entire sterile neutrino parameter space in order to see which parts can be ruled out with available data.

In Fig.~\ref{fig:constraints} we plot the parameter space in terms of mixing angle and mass (same as Fig.~\ref{fig:xmean}). Limits from X-ray observations by {\it Suzaku} are included as a thick black solid line (with grey shaded area). A more conservative limit based on Refs.~\cite{Canetti:2012kh,Ng:2015gfa} is added as a dashed line. The red point with error bar refers to the sterile neutrino interpretation of the suggested X-ray line signal~\cite{Bulbul:2014sua,Boyarsky:2014jta}.

The limit from Milky-Way satellite counts is illustrated as brown coloured area in Fig.~\ref{fig:constraints}. It excludes all models with particle mass below $m_{\rm sn}\sim 4.5$ keV and large parts of the parameter space above. The region around the claimed line signal is still allowed by satellite counts (as already shown by Refs.~\cite{Abazajian:2014gza, Merle:2014xpa, Lovell:2015psz}). Intriguingly, sterile neutrino DM with $m_{\rm sn}=7.1$ keV and a mixing angle of $\sin^2(2\theta)\sim 2-20\times 10^{-11}$ also seems to alleviate both the {\it missing satellite} and the {\it too-big-to-fail} problems \cite{Abazajian:2014gza,Bozek:2015bdo,Horiuchi:2015qri}.

The generalised Lyman-$\alpha$ limit based on the V13 reference model (excluding thermal relic WDM with $m_{\rm WDM}\lesssim3.3$ keV) is given by the green area in Fig.~\ref{fig:constraints}. It is considerably stronger than the bound from Milky-Way satellites and disfavours large parts of the remaining parameter space. Only a small area above $m_{\rm sn}\sim 10$ keV remains unchallenged by neither the V13 nor the X-ray bounds. Furthermore, the line signal at $m_{\rm sn}=7.1$ keV is in clear conflict with the V13 limits (as pointed out in Ref.~\cite{Merle:2014xpa}) putting further pressure on the sterile neutrino DM interpretation of the X-ray excess.

The generalised Lyman-$\alpha$ constraint based on the B15 reference model (excluding thermal relic WDM with $m_{\rm WDM}\lesssim4.35$ keV) is illustrated by the yellow area in Fig.~\ref{fig:constraints}. 
It completely overlaps with the {\it Suzaku} X-ray limits and therefore excludes the entire parameter space of resonantly produced sterile neutrino DM. Furthermore, the limit strongly disfavours the sterile neutrino interpretation of the suggested X-ray line signal.

In summary, the bounds presented in Fig.~\ref{fig:constraints} show for the first time that it is possible to not only rule out the non-resonant sterile neutrino scenario with structure formation, but to put strong pressure on the resonant production mechanism. However, before drawing final conclusions, it is important to note that the observational data used here could be subject to systematics which might somewhat reduce these limits. For example, the authors of Ref.~\cite{Garzilli:2015iwa} point out that the Lyman-$\alpha$ bounds of V13 could be relaxed to $m_{\rm WDM}\simeq 2.1$ keV, provided no specific temperature evolution of the intergalactic medium is assumed. This would shrink the green area to a size comparable to the brown area from satellite counts. On the other hand, B15 stress in their paper that the arguments of Ref.~\cite{Garzilli:2015iwa} do not affect their limits on the sterile neutrino particle mass (i.e the yellow area would not be reduced). The analysis of B15 could, however, suffer from systematics affecting the highest redshift bins of the BOSS data. B15 showed that ignoring these bins in their analysis significantly reduces the limit to $m_{\rm WDM}\sim 3.1$ keV (which would shrink the yellow area roughly down to the green one).

The bounds from dwarf galaxies could in principle be relaxed as well if the Milky-Way mass is considerably heavier or if the satellite distribution is more anisotropic than currently thought. On the other hand, the most recent discoveries of ultra-faint dwarf galaxies by DES seems to suggest the presence of many more nearly dark satellites below the detection limit of SDSS. Including these would tighten the constraints from dwarf galaxies, pushing the limits of the brown area in Fig.~\ref{fig:constraints} further to the right.

Finally, it is important to mention that the X-ray bounds depend on difficult flux measurements and should be interpreted with care. While there are several independent observations leading to similar (or even stronger) limits at masses below 10 keV, the {\it Suzaku} observations yield peerlessly tight bounds above this mass scale. For this reason, we also show independent X-ray bounds from \cite{Canetti:2012kh,Ng:2015gfa}, which still allow for resonant sterile neutrino DM with mass in the range of $m_{\rm sn}\sim15-30$ keV and mixing angles below $\sin^2(2\theta)\sim10^{-11}$.

Next to the observational uncertainties, there could also be systematics present in the theoretical calculation. The fact that sterile neutrino production is a resonant process which peaks during QCD transition makes theoretical predictions very challenging (as small errors could in principle significantly affect the result). In the past, there have been two main groups involved in the calculation of resonant sterile neutrino production, one based on initial work of Refs. \cite{Shi:1998km,Abazajian:2001nj} using a semi-classical Boltzmann approach and one applying a full field-theoretic calculation as presented in Ref. \cite{Asaka:2005an}. Unfortunately, no direct and systematic comparison between the two approaches currently exists. In the case of the 3.5 keV line signal, there are predictions from both approaches which agree reasonably well\footnote{A close inspection of the results from Ref.~ \cite{Lovell:2015psz}  and \cite{Venumadhav:2015pla} (based on the field theoretical and the Boltzmann approach, respectively) shows that the former yields slightly cooler power spectra than the latter. In terms of structure formation constraints shown in Fig.~\ref{fig:constraints}, this translates into a shift of order $\Delta m_{\rm sn}\sim 1$ keV to the left.}. Whether this is true for the entire parameter space remains an open question.

%%%%%%%%%%%%%%%%%%%%%%%%%
\section{\label{sec:conclusions}Conclusions}
The hypothetical sterile neutrino is an attractive dark matter (DM) candidate which only requires a minimal extension to the standard model of particle physics. As sterile neutrinos are not weakly interacting, they cannot be produced in the standard way by thermal freeze-out from the primordial plasma. The most popular production mechanism is by \emph{freeze-in} via resonantly enhanced mixing with the active neutrinos (i.e., the so-called Shi-Fuller mechanism).

Sterile neutrino DM consists of a testable scenario, expected to cause a line-signal in X-ray spectra (due to a decay channel to photons and active neutrinos) and to suppress structure formation at the smallest observable scales (due to significant free-streaming). These characteristics lead to independent constraints which have been used previously to rule out sterile neutrino DM based on non-resonant (Dodelson-Widrow) production \cite{Seljak:2006qw,Viel:2006kd,Horiuchi:2013noa}. The case of resonant production is more challenging because it yields non-standard momentum distributions affecting the details of small-scale clustering.

The aim of this paper is to constrain resonantly produced sterile neutrino DM with structure formation. We develop a method to generalise existing Lyman-$\alpha$ limits from thermal relic warm DM to the resonant sterile neutrino scenario. The method allows to exclude all resonant models which deviate more from the CDM baseline (at every data point of the Lyman-$\alpha$ flux power spectrum) than a previously excluded non-resonant reference model. Based on two reference models from Ref.~\cite{Baur:2015jsy} and \cite{Viel:2013apy} (referred to as B15 and V13, respectively), we are able to derive unprecedented bounds on the resonant scenario. Combined with the tightest bounds from X-ray observations, the former excludes the entire parameter space of resonant sterile neutrino DM at the 2-$\sigma$ confidence level, while the latter still leaves some room for sterile neutrinos with particle mass  above $m_{\rm sn}\sim 10$ keV and mixing angles below $\sin^22\theta\sim 5\times10^{-10}$.

In a next step, we apply the method developed in Ref.~\cite{Schneider:2014rda} to constrain resonant scenarios with Milky-Way satellite counts. We obtain less stringent limits than the ones from Lyman-$\alpha$, excluding all scenarios with particle mass below $m_{\rm sn}\sim 4.5$ keV.

Regarding the suggested X-ray line signal at $3.55$ keV \cite{Bulbul:2014sua,Boyarsky:2014jta}, we find the sterile neutrino interpretation to be clearly disfavoured by the limits from Lyman-$\alpha$ (based on both reference models) but consistent with the ones from Milky-Way satellite counts. This is in agreement with previous work from Ref.~\cite{Merle:2014xpa} and \cite{Horiuchi:2015qri,Lovell:2015psz}.

In summary, our findings show that sterile neutrino DM from resonant production is disfavoured by the combined limits from Lyman-$\alpha$, dwarf galaxy number counts, and X-ray data. However, we want to stress that these limits might be susceptible to theoretical and observational systematics which need to be fully understood before drawing final conclusions. Finally, it is worth mentioning that sterile neutrino DM based on production mechanisms which do not rely on the active-sterile mixing cannot be excluded by the constraints discussed above. This includes mechanisms based on the decay of the inflation \cite{Shaposhnikov:2006xi} or any other hypothetical scalar singlet \cite{Kusenko:2006rh,Petraki:2007gq,Merle:2013wta,Merle:2015oja}.

%%%%%%%%%%%%%%%%%%%%%%%%%

\section*{Acknowledgements}

I want to thank A. Merle and D. Reed for very helpful discussions. I am furthermore grateful to the authors of Ref.~\cite{Venumadhav:2015pla} for making public their code of resonant sterile neutrino production. Finally, I acknowledge support from the Swiss National Science Foundation via the Synergia project EUCLID.

%=============================================================================
\bibliographystyle{./apsrev}
\bibliography{ASbib}
%=============================================================================

\end{document}